\documentclass[journal]{IEEEtran}
\usepackage{amsmath}
\usepackage{amsfonts, amssymb}
\usepackage{algorithm}
\usepackage{algorithmic}
\usepackage{graphicx}
\usepackage{epstopdf}
\usepackage{subfigure}
\usepackage{booktabs}
\usepackage{threeparttable}
\usepackage{multirow}
\usepackage{tabularx}
\usepackage{color}
\usepackage{amsmath} 
\usepackage{amssymb}  
\usepackage{hyperref}
\DeclareMathOperator*{\argmin}{argmin}

\begin{document}

\title{Robust Matrix Completion via Maximum Correntropy Criterion and Half Quadratic Optimization\thanks{Yicong He, Fei Wang and Badong Chen are with the Institute of Artificial Intelligence and Robotics, Xi'an Jiaotong University, China, e-mails: heyicong@stu.xjtu.edu.cn, wfx@xjtu.edu.cn, chenbd@xjtu.edu.cn.} \thanks{ Yingsong Li is with the College of Information and Communication Engineering, Harbin Engineering University, Harbin 150001, China and National Space Science Center, Chinese Academy of Sciences, Beijing 100190, China, e-mail: liyingsong@ieee.org} \thanks{Jing Qin is with the Center of Smart Health, School of Nursing, The Hong
Kong Polytechnic University, Hongkong, China. e-mail:harryqinjingcn@gmail.com}}
%
\author{Yicong He, Fei Wang, \IEEEmembership{Member,~IEEE}, Yingsong Li, \IEEEmembership{Member,~IEEE}, Jing Qin, \IEEEmembership{Member,~IEEE} Badong Chen, \IEEEmembership{Senior~Member,~IEEE}}
%
%
%
%
\maketitle
\begin{abstract}
Robust matrix completion aims to recover a low-rank matrix from a subset of noisy entries perturbed by complex noises, where traditional methods for matrix completion may perform poorly due to utilizing $l_2$ error norm in optimization. In this paper, we propose a novel and fast robust matrix completion method based on maximum correntropy criterion (MCC). The correntropy based error measure is utilized instead of using $l_2$-based error norm to improve the robustness to noises. Using the half-quadratic optimization technique, the correntropy based optimization can be transformed to a weighted matrix factorization problem. Then, two efficient algorithms are derived, including alternating minimization based algorithm and alternating gradient descend based algorithm. The proposed algorithms do not need to calculate singular value decomposition (SVD) at each iteration. Further, the adaptive kernel selection strategy is proposed to accelerate the convergence speed as well as improve the performance. Comparison with existing robust matrix completion algorithms is provided by simulations, showing that the new methods can achieve better performance than existing state-of-the-art algorithms.
\end{abstract}
%
%
\section{Introduction}

\par Matrix completion is a novel technique that can fulfill the missing entries of a partially observed low-rank matrix \cite{Cand2009Exact,candes2010matrix,keshavan2010matrix}. Matrix completion takes advantages of low-rank properties of the matrix, which is always available in the fields such as computer vision \cite{Ji2010Robust,wu2010robust}, recommender systems \cite{koren2009matrix} and machine learning \cite{cabral2015matrix}, where the data can always be well approximated by low-rank representation or structure. A typical application of matrix completion is the recommender system \cite{isinkaye2015recommendation}. In the recommender system, how to guide users' behavior based on existing data is a crucial problem that should be considered. For example, Netflix--the world's largest online movie renter, which wants to recommend movies that might be of interest to users based on their behavior (ratings of various types of movies). User behavior consistency can be expressed by low rank property, thus matrix completion can be applied to predict the missing ratings of the users \cite{Netflix,zhou2008large}.

\par Although matrix completion is, in general, an NP-hard problem, in the last decades various algorithms have been proposed to tackle this problem and show accurate reconstruction performance. In particular, by formulating the problem as a constrained rank (or nuclear norm) minimization problem, several algorithms have been developed including normalized iterative hard thresholding (NIHT) \cite{Tanner2013Normalized}, iterative soft thresholding (IST) \cite{WRIGHT2009Sparse}, singular value thresholding (SVT) \cite{cai2010singular} and fixed point continuation (FPC) \cite{Ma2011Fixed}. These algorithms need to calculate full or truncated singular value decomposition (SVD) at each iteration, which may cause high computational complexity especially when the data scale is large. To reduce the performance degradation caused by SVD computation, recently the matrix factorization model is applied to solve the matrix completion problem \cite{keshavan2010matrix,Sun2015Guaranteed,hardt2014understanding,jain2013low}. In matrix factorization, the target matrix is represented as a multiple of two matrices so that the low-rank property can be guaranteed. Thus algorithms based on matrix factorization can naturally overcome the drawback of low efficiency in SVD computation. The representative algorithms include Power factorization (PF) \cite{haldar2009rank},  Low-rank Matrix Fitting (LMaFit) \cite{wen2012solving} and alternating steepest descent (ASD) \cite{tanner2016low}.

\par Traditional matrix completion often utilizes $l_2$ error norm in optimization, which can perform well under Gaussian noise assumption. However, in real applications, the observations may contain outliers. For example, in the recommender systems, the rate may exist human errors which is unreliable. In this case, the $l_2$ error norm may not properly represent the error statistics and the performance of traditional matrix completion algorithms may degrade. To improve the robustness against outliers, several robust matrix completion algorithms have been proposed. In \cite{Zeng2018Outlier}, the authors utilize $l_p$ error norm-based the cost function and solve the optimization problem by $l_p$ regression and alternating direction method of multipliers (ADMM). In \cite{Zhao2016Efficient}, the authors propose two new robust loss functions and utilize a distributed optimization framework \cite{scutari2014decomposition} to solve the problem in parallel.

\par In recent years, an information theoretic learning (ITL) \cite{principe2010information} measure called correntropy has been proposed to deal with the robust learning problem \cite{Zhao2011Kernel,Chen2015Maximum}. Correntropy is a smooth local similarity measure, which has its root in Renyi's entropy \cite{ITL}. With a Gaussian kernel, correntropy involves all the even moments of the error and is insensitive to large outliers \cite{liu2007correntropy}. Compared with $l_1$ norm, correntropy based methods can achieve better performance especially when the outliers are large \cite{Chen2016Generalized,ma2015maximum}. Applying correntropy to matrix completion, the correntropy based iterative soft and hard thresholding strategies have been proposed \cite{yang2018correntropy}. However, as mentioned before, the iterative thresholding based algorithms need to compute the SVD and will suffer from high computational cost when data scale is large.

\par In the present paper, we combine correntropy with matrix factorization method and propose a new cost function for robust matrix completion. The correntropy measure is utilized instead of using $l_2$ norm, thus the negative effects of outliers can be alleviated. Using matrix factorization, there is no need to compute the SVD at each iteration. Further, to efficiently solve the correntropy based optimization, the half-quadratic (HQ) technique is adopted \cite{nikolova2005analysis}. Using HQ, the complex optimization problem can be transformed into a quadratic optimization, and the traditional quadratic optimization method can be applied.

\par Based on HQ, we propose two algorithms for robust matrix completion. The first algorithm utilizes the traditional alternating minimization method \cite{jain2013low}. At each minimization step, the correntropy based optimization is transformed to a weighted least squares problem so that the solution can be iteratively obtained. The second algorithm directly transforms the correntropy based cost to a weighted matrix completion problem and then utilize the alternating steepest descend (ASD) method \cite{tanner2016low}. Both algorithms utilize HQ technique but in different ways for optimization. Moreover, taking advantage of the properties of correntropy, we propose an adaptive kernel width selection strategy for the proposed algorithms to improve the convergence speed as well as reconstruction accuracy. In summary, the main contributions of this paper are:
\par 1) A new cost function for robust matrix completion is proposed.
\par 2) Two efficient algorithms are developed via HQ techniques.
\par 3) Extensive simulations demonstrate the superior performance of the proposed algorithms compared with other state-of-the-art algorithms.

\par The paper is organized as follows. In section II we briefly review the concept of matrix completion and maximum correntropy criterion. In Section III we propose the new correntropy based matrix completion cost and propose two HQ based algorithms. In Section IV, simulation results are presented to demonstrate the reconstruction performance. Finally, conclusion is given in Section VI.

\section{Preliminaries}

\subsection{Matrix completion}
Consider a low-rank matrix $\boldsymbol{X}\in\mathbb{R}^{m\times n}$ where only a subset of entries can be observed. In particular, by defining the observed subset matrix $\boldsymbol{\Omega}\in\mathbb{R}^{m\times n}$ where
\begin{equation}
\boldsymbol{\Omega}_{i,j}=\left\{ {\begin{array}{*{20}{c}}
1,(i,j)\in{\Omega}\\
0,(i,j)\notin{\Omega}
\end{array}} \right.
\end{equation}
the observed matrix can be represented as $\boldsymbol{\Omega}\circ\boldsymbol{X}$, where $\circ$ denotes the Hadamard product. The goal of matrix completion is to recover the whole entries of $\boldsymbol{X}$ based on observed entries $\boldsymbol{\Omega}\circ\boldsymbol{X}$ and low rank property. In detail, the matrix completion can be formulated as the following constrained minimization problem
\begin{equation}
\label{NP}
\min_{\boldsymbol{M}} ~\rm{rank}(\boldsymbol{M})~~s.t.~~ \boldsymbol{\Omega}\circ\boldsymbol{M}=\boldsymbol{\Omega}\circ\boldsymbol{X}
\end{equation}
where $\boldsymbol{M}$ is the recovered matrix and $\circ$ denotes the Hadamard product.

The above optimization is an NP-hard and non-convex problem. In the last decade, various methods have been proposed to deal with the above matrix completion problem. Typically, there are three approaches, which are shown as follows:

1) Direct approach: Although Eq.(\ref{NP}) is NP-hard, methods based on iterative hard thresholding (IHT) technique \cite{blumensath2009iterative} can still be directly applied to the optimization problem. Similar to compressive sensing, at each iteration, the IHT approach utilizes gradient descent to decrease a measurement fidelity objective and then perform the best rank-R approximation. Note that the to obtain the largest R singular values at each iteration, the truncated SVD should be performed. Moreover, the normalized IHT (NIHT) \cite{Tanner2013Normalized} has also been introduced to matrix completion, which shows better performance than IHT.

2) Convex relaxation: A popular method for solving a non-convex optimization problem is to relax the nonconvex optimization to a convex problem. In particular, the convex nuclear norm is always used to replace the nonconvex rank minimization, i.e.
\begin{equation}
\label{convex}
\min_{\boldsymbol{M}} ~\|\boldsymbol{M}\|_{*}~~s.t.~~ \boldsymbol{\Omega}\circ\boldsymbol{M}=\boldsymbol{\Omega}\circ\boldsymbol{X}
\end{equation}
where $\|\boldsymbol{M}\|_{*}$ denotes the sum of all singular values of $\boldsymbol{M}$. The semidefinite programming (SDP) \cite{Liu2009Interior} and iterative soft thresholding (IST) \cite{WRIGHT2009Sparse} algorithms can be applied to solve Eq.(\ref{convex}) Note that to obtain the singular values, the SVD still should be performed at each iteration.

3) Matrix factorization: The above methods are both SVD based methods, and may suffer from high computational complexity when dealing with large scale data. Matrix factorization is a simple way to tackle this problem. Specifically, the recovered matrix $\boldsymbol{M}$ is factorized to multiple of two matrices $\boldsymbol{U}\in\mathbb{R}^{m\times r}$ and $\boldsymbol{V}\in\mathbb{R}^{r\times n}$ where $r$ is the rank of $\boldsymbol{M}$. The matrix factorization then solves the matrix completion by utilizing following objective function
\begin{equation}
\label{MF}
\min_{\boldsymbol{U},\boldsymbol{V}}{\|\boldsymbol{\Omega}\circ\left(\boldsymbol{X}-\boldsymbol{UV}\right)\|_{F}^2}
\end{equation}
where $\|\boldsymbol{X}\|_F$ denotes the Frobenius norm of matrix $\boldsymbol{X}$. The solution of (1.3) can be solved via alternating optimization methods. The representative algorithms include PowerFactorization (PF) \cite{haldar2009rank}, low-rank Matrix Fitting (LMaFit) \cite{wen2012solving} and alternating steepest descent (ASD) \cite{tanner2016low}.

\subsection{Maximum Correntropy Criterion}
Correntropy is a local and nonlinear similarity measure between two random variables within a "window" in the joint space determined by the kernel width.
Given two random variables $X$ and $Y$, the correntropy is defined by \cite{liu2007correntropy}
\begin{equation}
V (X, Y)= \boldsymbol{E}[\kappa (X, Y)]=\int \kappa (x, y)dF_{XY} (x, y)
\end{equation}
where $\kappa _\sigma$ is a shift-invariant Mercer kernel, and $F _{XY} (x, y)$ denotes the joint distribution function of $ (X, Y)$.
Given a finite number of samples $ \{x_i, y_i \} _{i=1}^N$, the correntropy can be approximated by
\begin{equation}
\hat{V} (X, Y)= \frac{1}{N} \sum_{i=1}^N \kappa (x_i, y_i)
\end{equation}
\par
In general, the kernel function of correntropy $\kappa (x, y)$ is the Gaussian kernel
\begin{equation}
\label{Gkernel}
\kappa (x, y)=G _\sigma (e) = \exp (-\frac{e^2}{2\sigma^2})
\end{equation}
where $e=x-y$ and $\sigma$ is the kernel width.
\par
\begin{figure}[tb]
\centering
\includegraphics[width=0.9\linewidth]{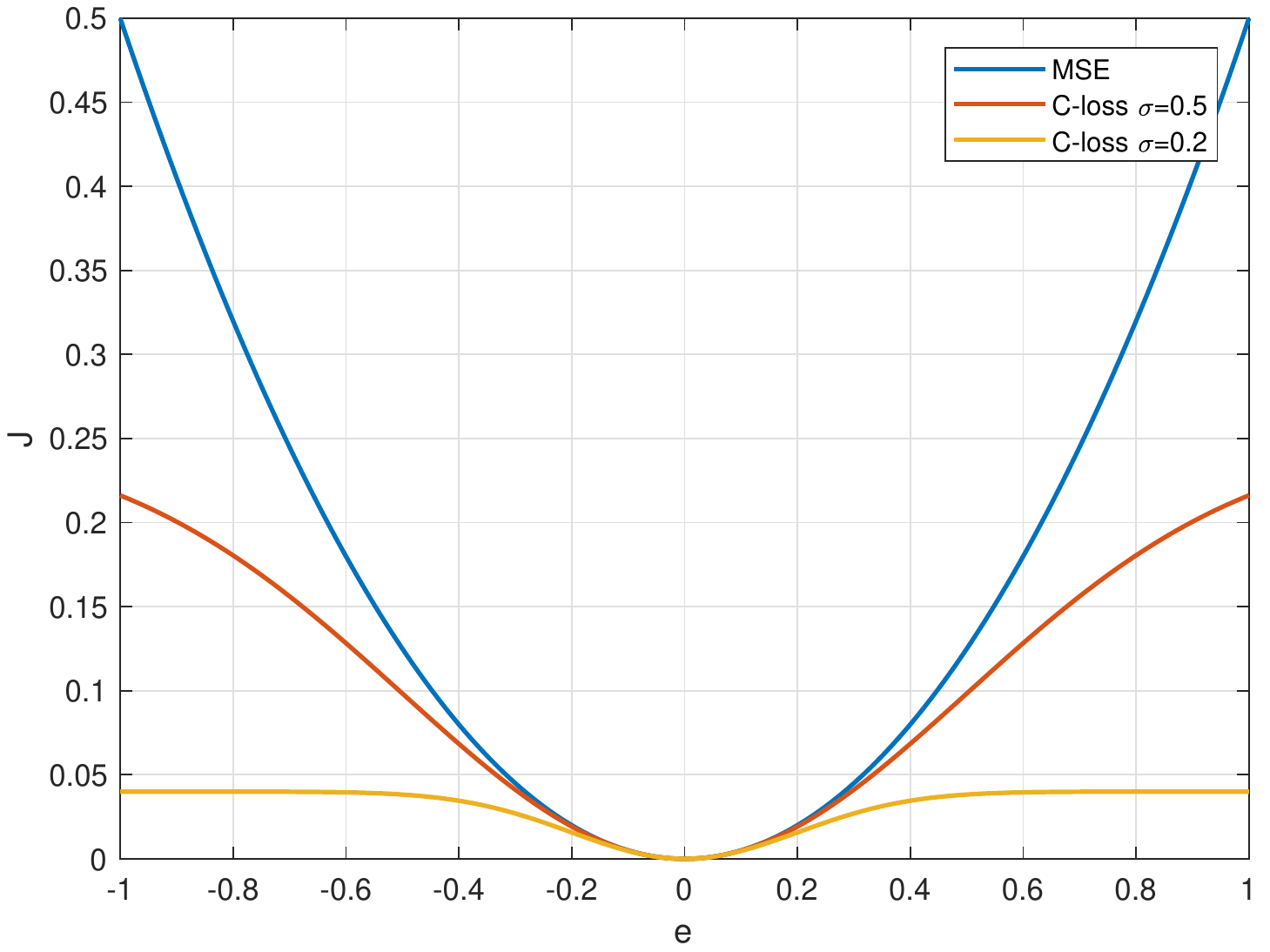}
\caption{Curves of ${J}$ versus error $e$ under different cost functions}
\label{fig1}
\end{figure}

Compared with the $l_2$ norm based second-order statistics of the error, correntropy involves all the even moments of the difference between $X$ and $Y$ and is insensitive to outliers. Replacing the second-order measure with correntropy measure leads to the maximum correntropy criterion (MCC) \cite{Singh2009Using}. The MCC solution is obtained by maximizing the following utility function
\begin{equation}
\label{MCC}
{J_{mcc}} = E\left[ {{G_{\sigma}\! }\left ( {{e (i)}} \right)}\right]
\end{equation}
Moreover, in practice, the MCC can also be formulated as minimizing the following correntropy-induced loss (C-loss) function \cite{Singh2014The,xu2018robust}
\begin{equation}
\label{C-loss}
J_{C-loss}= \frac{1}{M}\sum_{i=1}^M\sigma^2\left(1-{{G_{\sigma}\! }\left ( {{e (i)}} \right)}\right)
\end{equation}
The above cost function is closely related to the Welsch's cost function originally introduced in \cite{John1978Techniques}. The C-loss function with different kernel width $\sigma$ are shown in Fig.\ref{fig1}. One can observe that the C-loss function can effectively alleviate the impact of large errors. Moreover, selecting different kernel widths can adjust the range of sensitivity to outliers, while the error measure near zero will not be highly affected.

\section{Proposed Algorithms}

In this section, we combine matrix factorization with correntropy measure and propose a new objective function. Then we propose two efficient algorithms to solve the optimization problem.
\par Eq.(\ref{MF}) can be further rewritten as the sum of weighted squared residuals
\begin{equation}
\mathop {\min }\limits_{\boldsymbol{U},\boldsymbol{V}} J_2({\boldsymbol{U},\boldsymbol{V}})=\sum\limits_{i = 1}^n {\sum\limits_{j = 1}^m {\boldsymbol{\Omega}_{i,j}{{\left( {{\boldsymbol{X}_{i,j}} - {{\left( \boldsymbol{UV} \right)}_{i,j}}} \right)}^2}} }
\end{equation}
where $\boldsymbol{A}_{i,j}$ denotes the $\{i,j\}$-th entry of matrix $\boldsymbol{A}$. 
When the observed entry $\boldsymbol{X}_{i,j}$ contain large outliers, the $l_2$ error measure may not work well because the outliers may highly bias the optimization. To improve the robustness, in this paper we introduce the correntropy as the error measure. In particular, by replacing the $l_2$ error measure with correntropy, one can obtain the following new optimization problem:
\begin{equation}
\label{MFMCC}
\mathop {\min }\limits_{\boldsymbol{U},\boldsymbol{V}} J_{G_{\sigma}}({\boldsymbol{U},\boldsymbol{V}})\!=\!\sum\limits_{i = 1}^n \!{\sum\limits_{j = 1}^m \!\boldsymbol{\Omega}_{i,j}\sigma^2\left({1\!-\!G_{\sigma}{{\left( {{\boldsymbol{X}_{i,j}} \!-\! {{\left( \boldsymbol{UV} \right)}_{i,j}}} \right)}}} \right)}
\end{equation}
The above equation can also be simplified as the following representation
\begin{equation}
\label{MFMCCG}
\min_{\boldsymbol{U},\boldsymbol{V}}J_{G_{\sigma}}({\boldsymbol{U},\boldsymbol{V}})={\|\boldsymbol{\Omega}\circ\left(\boldsymbol{X}-\boldsymbol{UV}\right)\|_{G_{\sigma}}}
\end{equation}
The formulation of the above correntropy based optimization is closely related to \cite{he2011robust} and \cite{du2012robust}. In particular, when matrix $\boldsymbol{X}$ is fully observed (i.e. $\boldsymbol{\Omega}_{i,j}=1$ for all $i,j$), Eq.. equals to the optimization of robust PCA based on MCC \cite{he2011robust}. Further, when continues to impose constrains that both $\boldsymbol{U}$ and $\boldsymbol{V}$ are non-negative matrices, the optimization in Eq.(\ref{MFMCCG}) can be equivalent to the correntropy based nonnegative matrix factorization problem \cite{du2012robust}. Certainly, due to existence of observation matrix $\boldsymbol{\Omega}$ in Eq.(\ref{MFMCCG}), the solution in \cite{he2011robust} and \cite{du2012robust} will no longer suitable for matrix completion. Thus one should seek new approaches to solve Eq.(\ref{MFMCCG}).

\subsection{Optimization via half-quadratic}

In general, the correntropy based objective function in Eq.(\ref{MFMCCG}) is difficult to be minimized directly. To tackle this problem, the half-quadratic (HQ) technique has been applied to optimize these correntropy based cost functions \cite{he2011robust,du2012robust,he2014half,wang2017correntropy}. By introducing additional auxiliary variable, HQ reformulates a nonquadratic cost function as an augmented objective function in enlarged parameter space. 

According to half quadratic theory \cite{nikolova2007equivalence}, for $G _\sigma (e)$ there exists a convex conjugated function $\varphi$ so that
\begin{equation}
\label{HQ1}
G _\sigma (e)=\max_{t}\left(\frac{e^2t}{\sigma^2}-\varphi(t)\right)
\end{equation}
where $t\in\mathbb{R}$ and the maximum is reached at $t=-G _\sigma (e)$. Eq.(\ref{HQ1}) can be further rewritten as
\begin{equation}
\sigma^2(1-G _\sigma (e))=\min_{t}\left(-e^2t+\sigma^2\varphi(t)\right)
\end{equation}
By defining $s=-t$ and $\phi(s)=\sigma^2\varphi(-s)$, the above equation can be further derived as
\begin{equation}
\label{HQ2}
\min_{e}\sigma^2(1-G _\sigma (e))=\min_{e,s}\left(e^2s+\phi(s)\right)
\end{equation}
Thus, as shown above, minimizing the nonconvex C-loss function in terms of $e$ is equivalent to minimizing an augmented cost function in an enlarged parameter space $\{e,s\}$.
Therefore, by substituting Eq.(\ref{HQ2}) to Eq.(\ref{MFMCC}), the correntropy based objective function $J_{G_{\sigma}}({\boldsymbol{U},\boldsymbol{V}})$ can be further formulated as
\begin{equation}
\begin{aligned}
&J_{G_{\sigma}}({\boldsymbol{U},\boldsymbol{V}})\\
&=\mathop {\min }\limits_{\boldsymbol{W}} \sum\limits_{i = 1}^n \sum\limits_{j = 1}^m \left({{{W_{i,j}}{\Omega _{i,j}}{{\left( {{X_{i,j}} - {{\left( {UV} \right)}_{i,j}}} \right)}^2}} }  + {\Omega _{i,j}}\phi \left( {{W_{i,j}}} \right)\right)
\end{aligned}
\end{equation}
Further, by defining the augmented cost function
\begin{equation}
\label{HQ}
J_{HQ}({\boldsymbol{U},\boldsymbol{V}},\boldsymbol{W})={\|\sqrt{\boldsymbol{W}} \circ \boldsymbol{\Omega} \circ \left(\boldsymbol{X}-\boldsymbol{UV}\right)\|_{F}^2}+\boldsymbol{\Omega} \circ\phi \left( {\boldsymbol{W}}\right)
\end{equation}
where $\phi \left( {\boldsymbol{W}}\right)=\sum\nolimits_{i = 1}^n \sum\nolimits_{j = 1}^m \phi \left( {{W_{i,j}}} \right)$, we have the following relation
\begin{equation}
\label{HQG}
\min_{\boldsymbol{U},\boldsymbol{V}}J_{G_{\sigma}}({\boldsymbol{U},\boldsymbol{V}})=\min_{\boldsymbol{U},\boldsymbol{V},\boldsymbol{W}}J_{HQ}({\boldsymbol{U},\boldsymbol{V}},\boldsymbol{W})
\end{equation}
and the correntropy based optimization problem can be formulated as a half-quadratic based optimization. Similar to \cite{Zeng2018Outlier}, treating $\boldsymbol{U}$ and $\boldsymbol{V}$ as a whole, the optimization can be solved with following alternating minimization procedure:

1) optimizing $\boldsymbol{W}$: from Eq.(\ref{HQ1}) and Eq.(\ref{HQ2}) one can observe that given a certain $e$, the minimum is reached as $s=G_{\sigma}(e)$. Therefore, given the fixed $\boldsymbol{U}$ and $\boldsymbol{V}$, the optimal solutions of $W_{i,j}$ can be obtained as
\begin{equation}
\label{HQW}
W_{i,j}=G_{\sigma}{{\left( {{\boldsymbol{X}_{i,j}} - {{\left( \boldsymbol{UV} \right)}_{i,j}}} \right)}}, (i,j)\in\boldsymbol{\Omega}
\end{equation}
We should note that when $(i,j)\notin\boldsymbol{\Omega}$, optimal $W_{i,j}$ is unavailable. However, computing $W_{i,j}$ for $(i,j)\notin\boldsymbol{\Omega}$ do not affect solution of Eq.(\ref{MFMCCG}) and Eq.(\ref{HQ}) because of multiplication with $\boldsymbol{\Omega}$. To simplify the expression, in the following we directly use $W_{i,j}$ for full matrix ${\boldsymbol{W}}$. In practical application, one just need to calculate $W_{i,j}$ for $(i,j)\in\boldsymbol{\Omega}$.

2) Given a fixed $\boldsymbol{W}$, Eq.(\ref{HQ}) becomes a weighted matrix completion optimization problem
\begin{equation}
\label{W2}
\min_{\boldsymbol{U},\boldsymbol{V}}{\|\sqrt{\boldsymbol{W}} \circ \boldsymbol{\Omega} \circ \left(\boldsymbol{X}-\boldsymbol{UV}\right)\|_{F}^2}
\end{equation}
The weighting matrix $\boldsymbol{W}$ assigns different weights based on error residuals. According to the property of Gaussian function, a large error will assign small weight, such that the negative impact of outliers may be greatly reduced. So far, there are no existing algorithms to directly solve Eq.(\ref{HQ}). In the following, we follow two methods of traditional matrix completion and propose two efficient algorithms to solve the above weighted matrix completion optimization.

\subsection{Correntropy based power factorization algorithm}
In this part, we utilize the alternating minimization method to solve the correntropy based matrix completion problem. Alternating minimization is a widely used method for solving matrix factorization based optimization problem, and the representative algorithm is the Power factorization (PF). In particular, PF alternately minimizes $\boldsymbol{U}$ and $\boldsymbol{V}$ at each iteration for Eq.(\ref{MF}), i.e. fix one factored matrix and optimize the other. Similar to the traditional PF, for correntropy based optimization in Eq.(\ref{MFMCCG}), we can also alternatively optimize $\boldsymbol{U}$ and $\boldsymbol{V}$ as follows

\begin{equation}
\label{HQPFU}
\boldsymbol{U}^{t+1}=\argmin_{\boldsymbol{U}}{ \|\boldsymbol{\Omega} \circ \left(\boldsymbol{X}-\boldsymbol{U}\boldsymbol{V}^{t}\right)\|_{G_{\sigma}}}\\
\end{equation}
\begin{equation}
\label{HQPFV}
\boldsymbol{V}^{t+1}=\argmin_{\boldsymbol{V}}{ \|\boldsymbol{\Omega} \circ \left(\boldsymbol{X}-\boldsymbol{U}^{t+1}\boldsymbol{V}\right)\|_{G_{\sigma}}}\\
\end{equation}
where $t$ denotes the iteration number. Then, the HQ techniques can be utilized for each minimization step in Eq.(\ref{HQPFU}) or Eq.(\ref{HQPFV}).
In particular, similar to Eq.(\ref{HQG}), given a fixed $\boldsymbol{V}^t$
, Eq.(\ref{HQPFU}) can be further rewritten as %
\begin{equation}
\label{HQPFUG}
\boldsymbol{U}^{t+1}=\argmin_{\boldsymbol{U},\boldsymbol{W}}{\|\sqrt{\boldsymbol{W}} \circ \boldsymbol{\Omega} \circ \left(\boldsymbol{X}-\boldsymbol{UV}^t\right)\|_{F}^2}+\boldsymbol{\Omega} \circ\varphi \left( {\boldsymbol{W}}\right)
\end{equation}
The above minimization problem can be solved by alternating minimization described in the previous subsection. To distinguish the iteration procedure in Eq.(\ref{HQPFU}) and Eq.(\ref{HQPFV}), we denote the alternating minimization for $\boldsymbol{U}^{t+1}$ in Eq.(\ref{HQPFUG}) as the inner iteration, while the iteration in Eq.(\ref{HQPFU}) and Eq.(\ref{HQPFV}) are outer iteration. Therefore, at each inner iteration $k$, we first obtain optimal $\boldsymbol{W}$ from Eq.(\ref{HQW}) and then solve
\begin{equation}
\label{HQPFin}
\boldsymbol{U}^{k}=\argmin_{\boldsymbol{U}}{\|\sqrt{\boldsymbol{W}^k} \circ \boldsymbol{\Omega} \circ \left(\boldsymbol{X}-\boldsymbol{UV}\right)\|_{F}^2}
\end{equation}
Due to existence of $\boldsymbol{\Omega}$, Eq.(\ref{HQPFin}) do not has explicit solution. To solve this problem, similar to \cite{Zeng2018Outlier}, by defining
\begin{equation}
\begin{aligned}
\label{uxs}
&\boldsymbol{U}=[\boldsymbol{u}_1^T~\boldsymbol{u}_2^T~...~\boldsymbol{u}_m^T]^T, \boldsymbol{X}=[\boldsymbol{x}_1^T~\boldsymbol{x}_2^T~...~\boldsymbol{x}_m^T]^T\\ &\boldsymbol{S}=\sqrt{\boldsymbol{W}}\circ\boldsymbol{\Omega},\boldsymbol{S}=[\boldsymbol{s}_1^T~\boldsymbol{s}_2^T~...~\boldsymbol{s}_m^T]^T
\end{aligned}
\end{equation}
Eq.(\ref{HQPFin}) can be optimized by solving the following $m$ subproblems:
\begin{equation}
\boldsymbol{u}_i^k=\argmin_{\boldsymbol{u}}\left\|diag({\boldsymbol{s}^k_i}) \left(\boldsymbol{x}_i-\boldsymbol{V}^T\boldsymbol{u}\right)\right\|_2^2,~i=1,...,m
\end{equation}
$diag({\boldsymbol{s}_i^k})$ denotes the diagonal matrix whose entries are composed by elements of ${\boldsymbol{s}_i^k}$. For each subproblem, by dropping the zero diagonal entries of ${\boldsymbol{s}_i^k}$, one can finally obtain
\begin{equation}
\label{subproblem}
\boldsymbol{u}_i^{k}=\argmin_{\boldsymbol{u}}{\left\|{\sqrt{\boldsymbol{\Phi}^k}} \left(\boldsymbol{x}_{\boldsymbol{\theta}_i}-\boldsymbol{V}_{\boldsymbol{\theta}_i}^T\boldsymbol{u}\right)\right\|_{2}^2}
\end{equation}
where $\boldsymbol{\theta}_i$ denotes the index set of non-zero entries of ${\boldsymbol{s}^k_i}$, such that $\boldsymbol{x}_{\boldsymbol{\theta}_i}\in\mathbb{R}^{|\boldsymbol{s}_i^k|\times 1}$ and $\boldsymbol{V}_{\boldsymbol{\theta}_i}\in\mathbb{R}^{r \times |\boldsymbol{s}_i^k|}$ where $|\boldsymbol{s}_i^k|$ is the cardinality of ${\boldsymbol{s}_i^k}$. $\boldsymbol{\Phi}^k$ is a diagonal matrix with entries 
\begin{equation}
\label{phijj}
\boldsymbol{\Phi}_{j,j}^k=G_{\sigma}\left(\left(\boldsymbol{x}_{\boldsymbol{\theta}_i}-\boldsymbol{V}_{\boldsymbol{\theta}_i}^T\boldsymbol{u}\right)_j\right), j=1,2,...,|\boldsymbol{s}_i^k|\\
\end{equation}
Eq.(\ref{subproblem}) is essentially a weighted least squares problem and has a explicit solution
\begin{equation}
\label{ui}
\boldsymbol{u}_i^{k}=\left(\boldsymbol{V}_{\boldsymbol{\theta}_i}^T\boldsymbol{\Phi}^k\boldsymbol{V}_{\boldsymbol{\theta}_i}\right)^{-1}\boldsymbol{V}_{\boldsymbol{\theta}_i}^T\boldsymbol{\Phi}^k\boldsymbol{x}_{\boldsymbol{\theta}_i}
\end{equation}
Thus each $\boldsymbol{u}_i$ can be obtained by alternating calculate Eq.(\ref{phijj}) and Eq.(\ref{ui}) until convergence. The same iteration procedure can be applied for Eq.(\ref{HQPFV}) with fixed $\boldsymbol{U}$.

The above algorithm is called the half-quadratic based powerfactorization (HQ-PF) algorithm. In the following we give two propositions for HQ-PF.

Proposition 1: For a non-increasing $\sigma$, the sequence $\{J_{G_\sigma}({\boldsymbol{U}^t,\boldsymbol{V}^t}),t=1,2,...\}$ generated by HQ-PF will converge.

Proof: According to properties of alternating minimization, for a fixed $\sigma$ we can obtain
\begin{equation}
J_{G_\sigma}({\boldsymbol{U}^{t+1},\boldsymbol{V}^{t+1}})\leq J_{G_\sigma}({\boldsymbol{U}^{t+1},\boldsymbol{V}^{t}})\leq J_{G_\sigma}({\boldsymbol{U}^{t},\boldsymbol{V}^{t}})
\end{equation}
Moreover, considering the following function in terms of $\sigma$
\begin{equation}
f(\sigma)=\sigma^2(1-G_\sigma(x))
\end{equation}
Taking derivative of Eq.. we can obtain
\begin{equation}
\frac{\partial f(\sigma)}{\partial \sigma}=2\sigma(1-(1+\frac{x^2}{2\sigma^2})\exp(-\frac{x^2}{2\sigma^2}))\mathop  > \limits^{(a)}0
\end{equation}
The (a) holds since $e^y>y+1$ for $y>0$. Therefore $f(\sigma)$ monotonically increases as $\sigma$ increases, and then for $\sigma_1<\sigma_2$
$$J_{G_{\sigma_1}}({\boldsymbol{U}^{t+1},\boldsymbol{V}^{t+1}})\leq J_{G_{\sigma_2}}({\boldsymbol{U}^{t+1},\boldsymbol{V}^{t+1}})\leq J_{G_{\sigma_2}}({\boldsymbol{U}^{t},\boldsymbol{V}^{t}})$$
will always satisfied. Thus, for a non-increasing $\sigma$, the sequence $\{J_{G_\sigma}({\boldsymbol{U}^t,\boldsymbol{V}^t}),t=1,2,...\}$ monotonically non-increases
. It can be also verified that $J_{G_\sigma}({\boldsymbol{U}^t,\boldsymbol{V}^t})$ is always below-bounded for arbitrary $t$. Thus, $\{J_{G_\sigma}({\boldsymbol{U}^t,\boldsymbol{V}^t}),t=1,2,...\}$ will converge.

Proposition 2: When $\sigma\rightarrow\infty$, the HQ-PF is equal to PF.

Proof:
one can observe that when the kernel width $\sigma$ tends to infinity, the equation
\begin{equation}
\mathop{\lim} \limits_{\sigma\rightarrow+\infty}-2\sigma^2\exp\left (\! { - \frac{x^2}{2\sigma^2  \!}} \right)+ 2\sigma^2 = x^2
\end{equation}
will hold. Therefore, for sufficient large $\sigma$, the correntropy based optimization in Eq.(\ref{HQPFU}) and Eq.(\ref{HQPFV}) becomes equal to $l_2$ error norm based optimization
\begin{equation}
\boldsymbol{U}^{t+1}=\argmin_{\boldsymbol{U}}{ \|\boldsymbol{\Omega} \circ \left(\boldsymbol{X}-\boldsymbol{U}\boldsymbol{V}^{t}\right)\|_F^2}\\
\end{equation}
\begin{equation}
\boldsymbol{V}^{t+1}=\argmin_{\boldsymbol{V}}{ \|\boldsymbol{\Omega} \circ \left(\boldsymbol{X}-\boldsymbol{U}^{t+1}\boldsymbol{V}\right)\|_F^2}\\
\end{equation}
which is the typical iteration procedure of PF. Moreover, as $\sigma\rightarrow\infty$, $G_{\sigma}(e)$ will be equal to 1, and $\boldsymbol{\Phi}^k$ becomes the identity matrix. The $\boldsymbol{u}_i,i=1,...,m$ and $\boldsymbol{v}_j,j=1,...,n$ can be directly obtained by
\begin{equation}
\label{PF}
\boldsymbol{u}_i=\left(\boldsymbol{V}_{\boldsymbol{\theta}_i}^T\boldsymbol{V}_{\boldsymbol{\theta}_i}\right)^{-1}\boldsymbol{V}_{\boldsymbol{\theta}_i}^T\boldsymbol{x}_{\boldsymbol{\theta}_i}
\end{equation}
\begin{equation}
\label{PFV}
\boldsymbol{v}_j=\left(\boldsymbol{U}_{\boldsymbol{\theta}_j}^T\boldsymbol{U}_{\boldsymbol{\theta}_j}\right)^{-1}\boldsymbol{U}_{\boldsymbol{\theta}_j}^T\boldsymbol{x}_{\boldsymbol{\theta}_j}
\end{equation}
and the inner iteration number becomes 1. The solution coincides with the solution of PF \cite{haldar2009rank}.

As shown in Fig.\ref{fig1}, the kernel width $\sigma$ of Gaussian kernel function affects the range of sensitivity to outliers. Lots of works of literature have shown that relatively small $\sigma$ can offer more accurate performance but also suffers from low convergence speed \cite{huang2017adaptive}. A practical way is to use adaptive kernel width \cite{huang2017adaptive,ITL}. On the other hand, in the field of online adaptive filtering, many algorithms utilize the LS method at the first several iterations to speed up the convergence. In this work, to improve both the efficiency and accuracy, we combine the above two methods and propose a new kernel width selection strategy for HQPF. 
In particular, by defining the error residual matrix $\boldsymbol{E}^t$ at iteration $t$ with
\begin{equation}
\boldsymbol{E}^t=\boldsymbol{X}-\boldsymbol{U}^{t}\boldsymbol{V}^{t}
\end{equation}
we measure the convergence speed using the relative change of $\|\boldsymbol{E}^t\|_F$, i.e. $\|\boldsymbol{E}^{t+1}\|_F-\|\boldsymbol{E}^{t}\|_F$.
At the initial iterations, we directly utilize $l_2$ norm based PF solution in Eq.(\ref{PF}) to update $\boldsymbol{U}$ and $\boldsymbol{V}$ to speed up the initial convergence speed. When the convergence speed is slow, i.e.
$\|\boldsymbol{E}^{t+1}\|_F-\|\boldsymbol{E}^{t}\|_F<\varepsilon_1$
where $\varepsilon_1$ is the free threshold parameter, the optimization is switched to correntropy based optimization in Eq.(\ref{HQPFU}) and Eq.(\ref{HQPFV}). The adaptive kenrel width $\sigma^t$ at $t$-th iteration is selected using the following strategy
\begin{equation}
\label{sigmaout}
\sigma^{t}=\max(\eta\left({\boldsymbol{e}_{\Omega}^t}_{(0.75)}-{\boldsymbol{e}_{\Omega}^t}_{(0.25)}\right),\xi)
\end{equation}
where $\boldsymbol{e}_{\Omega}\in\mathbb{R}^{|\Omega|\times 1}$ denotes the vector composed by elements of all non-zero entries of $\boldsymbol{E}^t$, and $\boldsymbol{y}_{(q)}$ denotes the $q$-th quantile of $\boldsymbol{y}$. $\eta$ is the parameter controls the kernel width, and $\xi$ is the lower bound of $\sigma$. Finally, if  $\|\boldsymbol{E}^{t+1}\|_F-\|\boldsymbol{E}^{t}\|_F$ is less than a sufficient small value $\varepsilon_2$, we consider the algorithms converge to a local minimum, and the iteration procedure terminates.

For optimization of Eq.(\ref{HQPFin}), the selection of kernel width $\sigma_{in}$ for inner iteration also affects the performance. Too small $\sigma_{in}$ at initial iteration may lead to near-zero $\boldsymbol{W}$, and may cause the singularity problem. Therefore, we also utilize the adaptive kernel selection method to update $\sigma_{in}$ at each inner iteration. In particular, $\sigma_{in}^k$ is initialized to a sufficient large value (i.e.,10000) at $k=1$ and then update as follows for $k>1$:
\begin{equation}
\sigma_{in}^{k}=\left\{ {\begin{array}{*{20}{c}}
\left\|\frac{1}{2|s_i|} \left(\boldsymbol{x}_{s_i}-\boldsymbol{V}_{s_i}^T\boldsymbol{u}_i^k\right)\right\|_{2}&, \|\boldsymbol{u}_i^{k}-\boldsymbol{u}_i^{k-1}\|^2>\epsilon\\
\sigma^{t}&, \|\boldsymbol{u}_i^{k}-\boldsymbol{u}_i^{k-1}\|^2<\epsilon
\end{array}} \right.
\end{equation}
where $\epsilon$ is the threshold parameter. The norm of relative error vector $\|\boldsymbol{u}_i^{k}-\boldsymbol{u}_i^{k-1}\|^2$ is also utilized as the stop criterion for inner iteration.

The pseudocode of HQPF algorithm with adaptive kernel selection is summarized in Algorithm 1. Note that in each alternating minimization step, $m$ (or $n$) subproblems are actually independent with each other. Thus one can further utilize a distributed system to solve the subproblems in parallel and speed up the computation.
\begin{algorithm}
\caption{HQ-PF for robust matrix completion}
\begin{algorithmic}
 \REQUIRE $\boldsymbol{\Omega}$, $\boldsymbol{\Omega}\circ\boldsymbol{X}$ and r
 \STATE \emph{\%Initialization}
 \STATE initial matrices $\boldsymbol{U}^0$ and $\boldsymbol{V}^0$, $\boldsymbol{E}^0=\boldsymbol{0}$, $t=0$\\
 \STATE \emph{\%Computation using $l_2$ norm based solution}
 \REPEAT
 \STATE solve $\boldsymbol{u}_i^{t+1},i=1,...,m$ using (\ref{PF})
 \STATE solve $\boldsymbol{v}_i^{t+1},i=1,...,n$ using (\ref{PFV})
 \STATE $t=t+1$
 \UNTIL $\|\boldsymbol{E}^{t}\|_F-\|\boldsymbol{E}^{t-1}\|_F<\varepsilon_1$
 \STATE \emph{\%Computation using correntropy based solution}
 \REPEAT
 \STATE compute $\sigma^t$ according to (\ref{sigmaout})
 \STATE solve $\boldsymbol{u}_i^{t+1},i=1,...,m$ alternatively computing (\ref{phijj}) and (\ref{ui}) until convergence
 \STATE solve $\boldsymbol{v}_i^{t+1},i=1,...,n$ using the same method with $\boldsymbol{u}_i$ until convergence
 \STATE $t=t+1$
 \UNTIL $\|\boldsymbol{E}^t\|_F-\|\boldsymbol{E}^{t-1}\|_F<\varepsilon_2$

\ENSURE $\boldsymbol{M}=\boldsymbol{\boldsymbol{U}}^{t}{\boldsymbol{V}}^{t}$
\end{algorithmic}
\end{algorithm}

\subsection{Correntropy based alternating steepest descent algorithm}

HQ-PF is an extension of the traditional PF algorithm. Although HQ-PF is a distributable algorithm which can improve computation efficiency, the whole computational cost is still much higher than $l_2$ based algorithm since at each iteration the weighted LS optimization should be applied. Recently, an alternating steepest descent (ASD) method is proposed for matrix completion task. ASD directly applies gradient descend method and shows faster performance than alternating minimization based algorithms. Inspired by ASD, in this section we introduce the gradient descent method to solve Eq.(\ref{MFMCCG}) and derive a more efficient algorithm.

As described in subsection A, we first optimize $\boldsymbol{W}$ according to Eq.(\ref{HQW}) Then, unlike alternating minimization, we directly apply the gradient descent method to alternative update $\boldsymbol{U}$ and $\boldsymbol{V}$ only one step at each iteration. For further derivation, we add a coefficient to Eq.(\ref{W2}) so that the minimization becomes
\begin{equation}
\label{W3}
\min_{\boldsymbol{U},\boldsymbol{V}}\frac{1}{2}{\|\sqrt{\boldsymbol{W}} \circ \boldsymbol{\Omega} \circ \left(\boldsymbol{X}-\boldsymbol{UV}\right)\|_{F}^2}
\end{equation}
then, based on Eq.(\ref{uxs}), for a fixed $\boldsymbol{V}$, Eq.(\ref{W3}) can be rewritten as the function in terms of $\boldsymbol{U}$:
\begin{equation}
\label{fV}
f_{\boldsymbol{V}}(\boldsymbol{U})=\frac{1}{2}\sum_{i=1}^{n}(\boldsymbol{x}_i-\boldsymbol{V}^T\boldsymbol{u}_i)\boldsymbol{\Sigma}_i(\boldsymbol{x}_i-\boldsymbol{V}^T\boldsymbol{u}_i)^T
\end{equation}
where
$\boldsymbol{\Sigma}_i=diag({\boldsymbol{s}_i^2})$.
Thus the gradient of Eq.(\ref{fV}) at each element $u_{ij}$ can be derived as
\begin{equation}
\begin{aligned}
\frac{\partial f_{\boldsymbol{V}}(\boldsymbol{U})}{\partial u_{ij}}&=-(\boldsymbol{x}_i\boldsymbol{\Sigma}_i\boldsymbol{V})_j+(\boldsymbol{u}_i\boldsymbol{V}\boldsymbol{\Sigma}_i\boldsymbol{V})_j\\
&=-(\boldsymbol{W}\circ\boldsymbol
{\Omega}\circ \boldsymbol{XV})_{ij}+(\boldsymbol{W}\circ\boldsymbol{\Omega}\circ (\boldsymbol{UV})\boldsymbol{V})_{ij}\\
&=-(\boldsymbol{W}\circ\boldsymbol{\Omega}\circ(\boldsymbol{X}-\boldsymbol{UV})\boldsymbol{V})_{ij}
\end{aligned}
\end{equation}
hence the gradient descent of $f_{\boldsymbol{V}}(\boldsymbol{U})$ along $\boldsymbol{U}$ can be obtained as
\begin{equation}
\label{gU}
\boldsymbol{g}_{\boldsymbol{U}}=\frac{\partial f_{\boldsymbol{V}}(\boldsymbol{U})}{\partial \boldsymbol{U}}=-\boldsymbol{W}\circ\boldsymbol{\Omega}\circ(\boldsymbol{X}-\boldsymbol{UV})\boldsymbol{V}^T
\end{equation}
Further, the gradient descent step size $\mu_{\boldsymbol{U}}$ is selected by solving the following optimization problem
\begin{equation}
\begin{aligned}
\label{muU}
\mu_{\boldsymbol{U}}&=\argmin_{\mu}{\|\sqrt{\boldsymbol{W}} \circ \boldsymbol{\Omega} \circ \left(\boldsymbol{X}-(\boldsymbol{U}-\mu\boldsymbol{g}_{\boldsymbol{U}})\boldsymbol{V}\right)\|_{F}^2}\\
&=\frac{\|\boldsymbol{g}_{\boldsymbol{U}}\|_F^2}{\|\sqrt{\boldsymbol{W}} \circ \boldsymbol{\Omega} \circ \left(\boldsymbol{g}_{\boldsymbol{U}}\boldsymbol{V}\right)\|_F^2}
\end{aligned}
\end{equation}
Similar to Eq.(\ref{gU}) and Eq.(\ref{muU}), for a fixed $\boldsymbol{U}$, we can obtain the gradient descent along $\boldsymbol{V}$ and the corresponding step size as
\begin{equation}
\label{gV}
\boldsymbol{g}_{\boldsymbol{V}}=-\boldsymbol{U}^T (\boldsymbol{W}\circ\boldsymbol{\Omega}\circ (\boldsymbol{X}-\boldsymbol{UV}))
\end{equation}
\begin{equation}
\label{muv}
\mu_{\boldsymbol{V}}=\frac{\|\boldsymbol{g}_{\boldsymbol{V}}\|_F^2}{\|\sqrt{\boldsymbol{W}} \circ \boldsymbol{\Omega} \circ \left(\boldsymbol{U}\boldsymbol{g}_{\boldsymbol{V}}\right)\|_F^2}
\end{equation}
Therefore, the matrices $\boldsymbol{U}$ and $\boldsymbol{V}$ can be alternated update using gradient descend method, i.e. for each iteration $t$.
\begin{equation}
\begin{aligned}
\boldsymbol{U}^{t+1}=\boldsymbol{U}^t-\mu^{t}_{\boldsymbol{U}}\boldsymbol{g}_U^t\\
\boldsymbol{V}^{t+1}=\boldsymbol{V}^t-\mu^{t}_{\boldsymbol{V}}\boldsymbol{g}_V^t
\end{aligned}
\end{equation}

The algorithm with update above is called the half-quadratic alternating steepest descend (HQASD) algorithm. The following proposition guarantees the convergence of the $J_{G_\sigma}({\boldsymbol{U},\boldsymbol{V}})$ using the above gradient descend method.

Proposition 3: For a non-increasing $\sigma$, the sequence $\{J_{G_\sigma}({\boldsymbol{U}^t,\boldsymbol{V}^t}),t=1,2,...\}$ generated by HQASD will converge.

Proof: according to properties of alternating descend, for a fixed $\sigma$ we can obtain
\begin{equation}
\begin{aligned}
J_{HQ}({\boldsymbol{U}^{t+1},\boldsymbol{V}^{t+1},\boldsymbol{W}^{t+1}})&\leq J({\boldsymbol{U}^{t+1},\boldsymbol{V}^{t+1},\boldsymbol{W}^{t}})\\
&\leq J_{HQ}({\boldsymbol{U}^{t+1},\boldsymbol{V}^{t},\boldsymbol{W}^{t}})\\
&\leq J_{HQ}({\boldsymbol{U}^{t},\boldsymbol{V}^{t},\boldsymbol{W}^{t}})
\end{aligned}
\end{equation}
Since $J_{HQ}$ is bounded below, one can obtain \cite{nikolova2005analysis}
$$J_{G_\sigma}({\boldsymbol{U}^{t+1},\boldsymbol{V}^{t+1}})\leq J_{G_\sigma}({\boldsymbol{U}^{t},\boldsymbol{V}^{t}})$$
Moreover, according to proof of proposition 1, for $\sigma_1<\sigma_2$
$$J_{G_{\sigma_1}}({\boldsymbol{U}^{t+1},\boldsymbol{V}^{t+1}})\leq J_{G_{\sigma_2}}({\boldsymbol{U}^{t},\boldsymbol{V}^{t}})$$
will always satisfied. Thus, the sequence $\{J_{G_\sigma}({\boldsymbol{U}^t,\boldsymbol{V}^t}),t=1,2,...\}$ generated by HQ-ASD will converge for non-increasing $\sigma$ .

Proposition 4: When $\sigma\rightarrow\infty$, the HQ-PF is equal to PF.

Proof:
As $\sigma\rightarrow\infty$, $G_{\sigma}(e)$ will be equal to 1, thus all the entries of $\boldsymbol{W}$ becomes 1. The $\boldsymbol{W}$ in Eq.(\ref{HQW}) does not need to be optimized, and Eq.(\ref{gU})-(\ref{muv}) will be the same as the algorithm for ASD in \cite{tanner2016low}.

In \cite{tanner2016low}, the author also proposed the scaled ASD (ScaledASD) algorithm to improve the convergence speed and recover performance. Similar to ScaledASD, we can further scale the gradient descent directions in Eq.(\ref{gU}) and Eq.(\ref{gV}) by $({\boldsymbol{V}}{\boldsymbol{V}}^T)^{-1}$ and $({\boldsymbol{U}}^T{\boldsymbol{U}})^{-1}$, respectively, i.e.
\begin{equation}
\begin{aligned}
\hat{\boldsymbol{g}}_{\boldsymbol{U}}=\frac{\boldsymbol{g}_U}{({\boldsymbol{V}}{\boldsymbol{V}}^T)^{-1}}\\
\hat{\boldsymbol{g}}_{\boldsymbol{V}}=\frac{\boldsymbol{g}_V}{({\boldsymbol{U}}^T{\boldsymbol{U}})^{-1}}
\end{aligned}
\end{equation}
the corresponding step sizes are obtained as
\begin{equation}
\begin{aligned}
\hat{\mu}_{\boldsymbol{U}}=\frac{tr\{\boldsymbol{g}_{\boldsymbol{U}}^T\hat{\boldsymbol{g}}_{\boldsymbol{U}}\}}{\|\sqrt{\boldsymbol{W}} \circ \boldsymbol{\Omega} \circ \left(\hat{\boldsymbol{g}}_{\boldsymbol{U}}\boldsymbol{V}\right)\|_F^2}\\
\hat{\mu}_{\boldsymbol{V}}=\frac{tr\{\boldsymbol{g}_{\boldsymbol{V}}^T\hat{\boldsymbol{g}}_{\boldsymbol{V}}\}}{\|\sqrt{\boldsymbol{W}} \circ \boldsymbol{\Omega} \circ \left(\boldsymbol{U}\hat{\boldsymbol{g}}_{\boldsymbol{V}}\right)\|_F^2}
\end{aligned}
\end{equation}
and the gradient update of $\boldsymbol{U}$ and $\boldsymbol{V}$ can be then derived as
\begin{equation}
\begin{aligned}
\label{scaledASD}
\boldsymbol{U}^{t+1}=\boldsymbol{U}^t-\hat{\mu}^{t}_{\boldsymbol{U}}\hat{\boldsymbol{g}}_U^t\\
\boldsymbol{V}^{t+1}=\boldsymbol{V}^t-\hat{\mu}^{t}_{\boldsymbol{V}}\hat{\boldsymbol{g}}_V^t
\end{aligned}
\end{equation}
Since ScaledASD has been proved to show better performance than ASD \cite{tanner2016low}. Thus, we can conduct that Scaled HQ-ASD will also perform better than Scaled HQ-ASD. Therefore, for simplicity, in the following part we directly utilize Eq.(\ref{scaledASD}) as the update of HQ-ASD.

Similar to HQPF, we also apply the adaptive selection of kernel width $\sigma$ to improve the convergence speed and performance of HQASD. In particular, at first several iterations, the kernel width $\sigma^t$ is fixed to sufficient large value (or equivalently set $\boldsymbol{W}$ to an all one matrix and use ASD update procedure). When $\|\boldsymbol{E}^{t+1}\|_F-\|\boldsymbol{E}^{t}\|_F<\varepsilon_3$
, the optimization is switched to correntropy based optimization and the HQASD with adaptive kernel width in Eq.(\ref{sigmaout}) is applied.

The pseudocode of HQASD is summarized in Algorithm 2.
\begin{algorithm}
\caption{HQASD for robust matrix completion}
\begin{algorithmic}
 \REQUIRE $\boldsymbol{\Omega}$, $\boldsymbol{\Omega}\circ\boldsymbol{X}$ and r
 \STATE \emph{\%Initialization}
 \STATE initial matrices $\boldsymbol{U}^0$ and $\boldsymbol{V}^0$, $\boldsymbol{E}^0=\boldsymbol{0}$, $t=0$, $\sigma^0=10000$\\
 \STATE \emph{\%Computation using ASD (i.e. sufficient large $\sigma$)}
 \REPEAT
 \STATE $\sigma^t=10000$
 \STATE compute $\boldsymbol{U}^{t+1},\boldsymbol{V}^{t+1}$ using (\ref{scaledASD})
 \STATE $t=t+1$
 \UNTIL $\|\boldsymbol{E}^{t}\|_F-\|\boldsymbol{E}^{t-1}\|_F<\varepsilon_3$
 \STATE \emph{\%Computation using correntropy based solution}
 \REPEAT
 \STATE compute $\sigma^t$ according to (\ref{sigmaout})
 \STATE compute $\boldsymbol{U}^{t+1},\boldsymbol{V}^{t+1}$ using (\ref{scaledASD})
 \STATE $t=t+1$
 \UNTIL $\|\boldsymbol{E}^t\|_F-\|\boldsymbol{E}^{t-1}\|_F<\varepsilon_4$

\ENSURE $\boldsymbol{M}=\boldsymbol{\boldsymbol{U}}^{t}{\boldsymbol{V}}^{t}$
\end{algorithmic}
\end{algorithm}

\subsection{Complexity analysis}
In this part, we discuss the complexity of the two proposed algorithms. For HQPF, at each minimization step of Eq.({HQPFU}) and Eq.(\ref{HQPFV}), the complexity is $o(|\Omega|r^2N_{HQ})$ where $N_{HQ}$ is the number of outer iteration. For the inner iteration of HQPF, we consider two cases. When utilizing PF at first several iterations (denoted as $K_2$), the least squares solution can be directly obtained, such that $N_{HQ}=1$. When applied weighted least squares for HQPF, an iteration procedure should be performed, and the inner iteration number is denoted as $K_{HQ}$. Therefore, the final complexity of HQPF is $o(|\Omega|r^2(K_2+N_{HQ}K_{HQ}))$. One can observe that the complexity is closely related to the percentage of observations and rank of $r$. Larger rank or larger amount of observed entries may both increase the computational cost of HQPF. Moreover, as mentioned in Section B, HQPF is friendly to multicore and distributed systems. In particular, subproblems for solving $\boldsymbol{u}_i,i=1,...,m$ and $\boldsymbol{v}_i,i=1,...,n$ are independent and can be applied in a parallel way. Therefore, for a distributed system with $p$ workers, the complexity of each worker will be reduced to $o(|\Omega|r^2(K_2+N_{HQ}K_{MCC})/p)$.

The complexity of HQASD is similar to ASD \cite{tanner2016low}. In particular, the complexity per iteration without $\boldsymbol{W}$ can be directly obtained from the complexity of ASD, i.e. the complexity is $o(|\Omega|r)$. When taking computation of $W$ into consideration, the complexity of MCC-ASD at each iteration becomes $o(|\Omega|r+1)$. Therefore, the overall complexity of MCC-ASD is $o((|\Omega|r+1)K_{ASD})$ where $K_{ASD}$ is the iteration number. As can be seen, the complexity of HQASD is also positively correlated to the percentage of observations and rank of $\boldsymbol{X}$. Moreover, compared with HQPF, the complexity of HQASD per iteration is much smaller especially when rank $r$ or matrix size is large (large matrix size will lead to large $N_{HQ}$). Certainly, the gradient descend based HQASD may need a larger number of iterations than HQPF. The final computation cost comparison will be verified by simulations.

\section{Simulations}
\label{sec:pagestyle}

In this section, we carry out simulations to verify the performance of the proposed two algorithms.

We compare the performance with existing state-of-the-art robust matrix completion methods including $l_1$ based alternating minimization via PowerFactorization ($l_1$-PF), Quadratic Programming (QP) with the loss function $f(x)=1/\beta\cdot\log((e^{\beta x}+e^{-\beta x})/2)$ (BMFC) and correntropy based iterative hard thresholding (CIHT). In particular, to ensure fairness in comparison, the kernel width adaptation method proposed in this paper is also applied to CIHT in the simulations. All the algorithms are implemented in MATLAB r2017b on a 2.6-GHz and 16-GB memory computer without any acceleration. The completion performance is evaluated by normalized mean square error (NMSE) defined by
\begin{equation}
\label{NMSE}
NMSE=\frac{{{E\left[ {\left\|{\hat {\boldsymbol{M}}}-\boldsymbol{X} \right\|}_F^2\right]}}}{{{\| \boldsymbol{X} \|}_F^2}}
\end{equation}
In the simulation, the expectation in Eq.(\ref{NMSE}) is approximated by
$${E\left[ {\left\|{\hat {\boldsymbol{M}}}-\boldsymbol{X} \right\|}_F^2\right]}\approx\frac{1}{T}\sum_{mc=1}^{T}{{\| {{\hat {\boldsymbol{M}}}^{mc}}-\boldsymbol{X} \|}_F^2}$$
where $T$ is the number of Monte Carlo runs.

In the simulation, the typical two-component Gaussian mixture model (GMM) is utilized as the non-Gaussian noise model. The probability density function (PDF) of GMM is defined as
\begin{equation}
p_v(i)=(1-c)N(0,{\sigma_A^2})+cN(0,\sigma_B^2)
\end{equation}
where $N(0,{\sigma_A^2})$ represents general noise disturbance with variance ${\sigma_A^2}$, and $N(0,\sigma_B^2)$ stands for outliers that occur occasionally with a large variance $\sigma_B^2$. The variable $c$ controls the occurrence probability of outliers.

For all algorithms, similar to HQPF and HQASD, we utilize the relative change of the current and previous iterations $\|\boldsymbol{E}^t\|_F-\|\boldsymbol{E}^{t-1}\|_F$ as the stop criterion, and the threshold parameter is set specific to each algorithm. During all the simulations, without explicitly mentioned, the threshold parameter $\varepsilon_2$ for stop criterion is set to $10e^{-3}$ for BMFC, $l_1$-PF, HQ-PF, and $10e^{-7}$ for HQASD and CIHT. The threshold parameter $\varepsilon_1$ for adaptive kernel width strategy is set to $10e^{-2}$, $10e^{-4}$ and $10e^{-3}$ for HQ-PF, HQASD and CIHT, respectively. The inner iteration threshold $\epsilon$ for weighted LS is set to $10e^{-8}$ for both $l_1$-PF and HQPF. Other parameters are tuned to achieve the best during each simulation.

\begin{figure}[tb]
\centering
\includegraphics[width=0.9\linewidth]{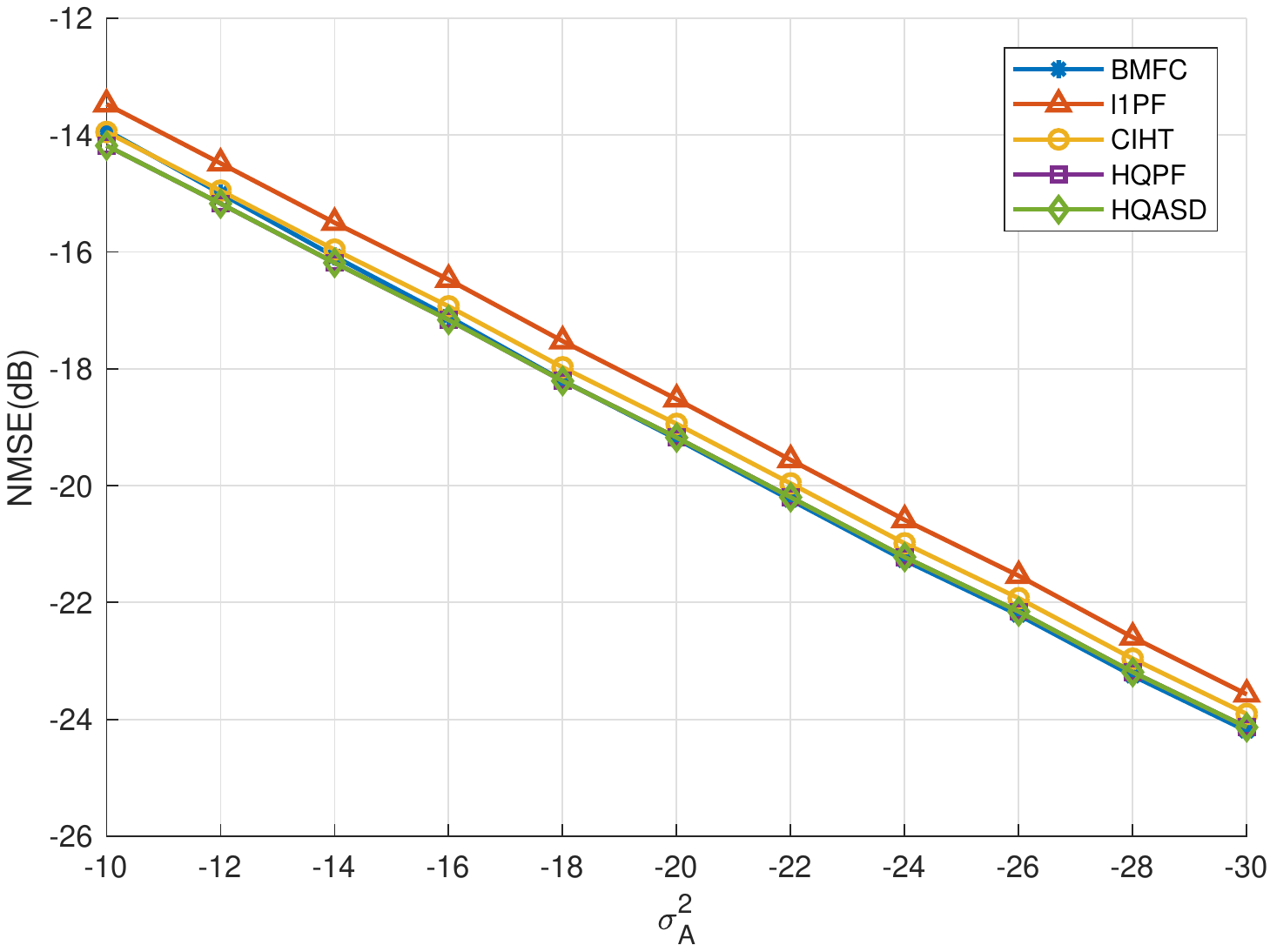}
\caption{Curves of NMSE with different $\sigma_{A}^2$(dB).}
\label{fig2}
\end{figure}

\begin{figure}[tb]
\centering
\includegraphics[width=0.9\linewidth]{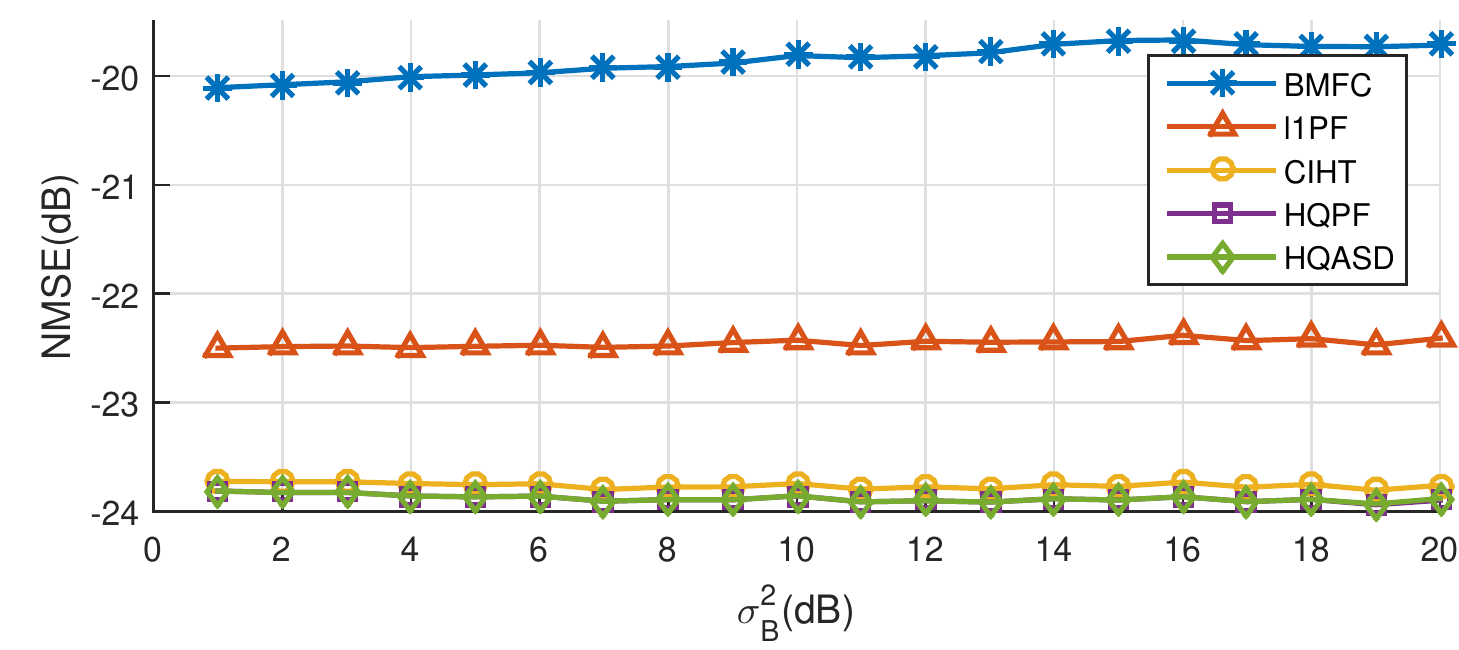}
\caption{Curves of NMSE with different $\sigma_{B}^2$(dB).}
\label{fig3}
\end{figure}

\begin{figure}[tb]
\centering
\includegraphics[width=0.9\linewidth]{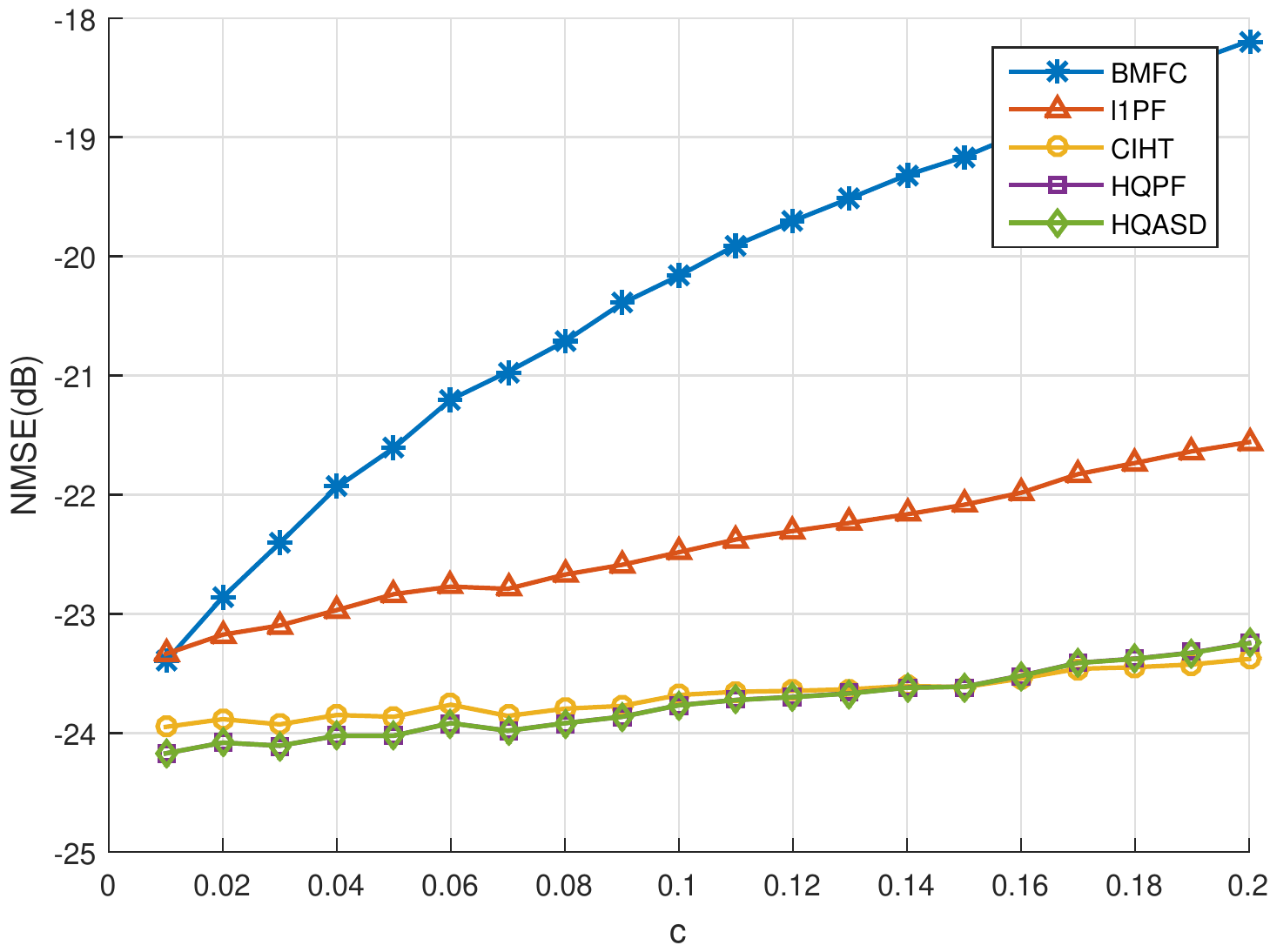}
\caption{Curves of NMSE with different $c$.}
\label{fig4}
\end{figure}

\subsection{Random matrix completion}
We first compare the performance on the synthetic random data. The matrix $\boldsymbol{X}$ with rank $r$ is generated by multiplying two matrices $\boldsymbol{U}$ and $\boldsymbol{V}$. The observation matrix $\boldsymbol{\Omega}$ with the percentage of observation $p$ is generated by randomly assign $p\%$ of the entries of $\boldsymbol{\Omega}$ with value 1. In this part, without explicitly mentioned, we set the matrix $\boldsymbol{X}$ as the squared matrix and the dimension is set to $m=n=256$. The parameters of GMM noise are set as $\sigma_A=0.01,\sigma_B=1,c=0.1$. The rank is set to $r=5$. The percentage of observation is fixed at $60\%$. The parameter $a$ for BMFC is set to 6. For each simulation, the NMSE is obtained via 200 Monte Carlo runs with different realizations of $\boldsymbol{X}$, $\boldsymbol{\Omega}$, $\boldsymbol{U}^0$, $\boldsymbol{V}^0$ and noises.
\par First, we compare the performance under different noise environments. We gradually increase the variance $\sigma_A^2$, $\sigma_B^2$, $c$ and calculate the NMSE values. The curves of NMSE with different noise distributions are shown in Fig.\ref{fig2}-\ref{fig4}. One can observe that the performance of all algorithms degrade as $\sigma_A^2$ and $c$ increases, while the performance is slightly changed with different levels of outliers (i.e. $\sigma_B^2$). In particular, all algorithms achieve comparable performance except $l_1$-PF, which is mainly caused by nonsmooth of $l_1$ norm near zero error.
\par
\begin{figure}[tb]
\centering
\includegraphics[width=0.9\linewidth]{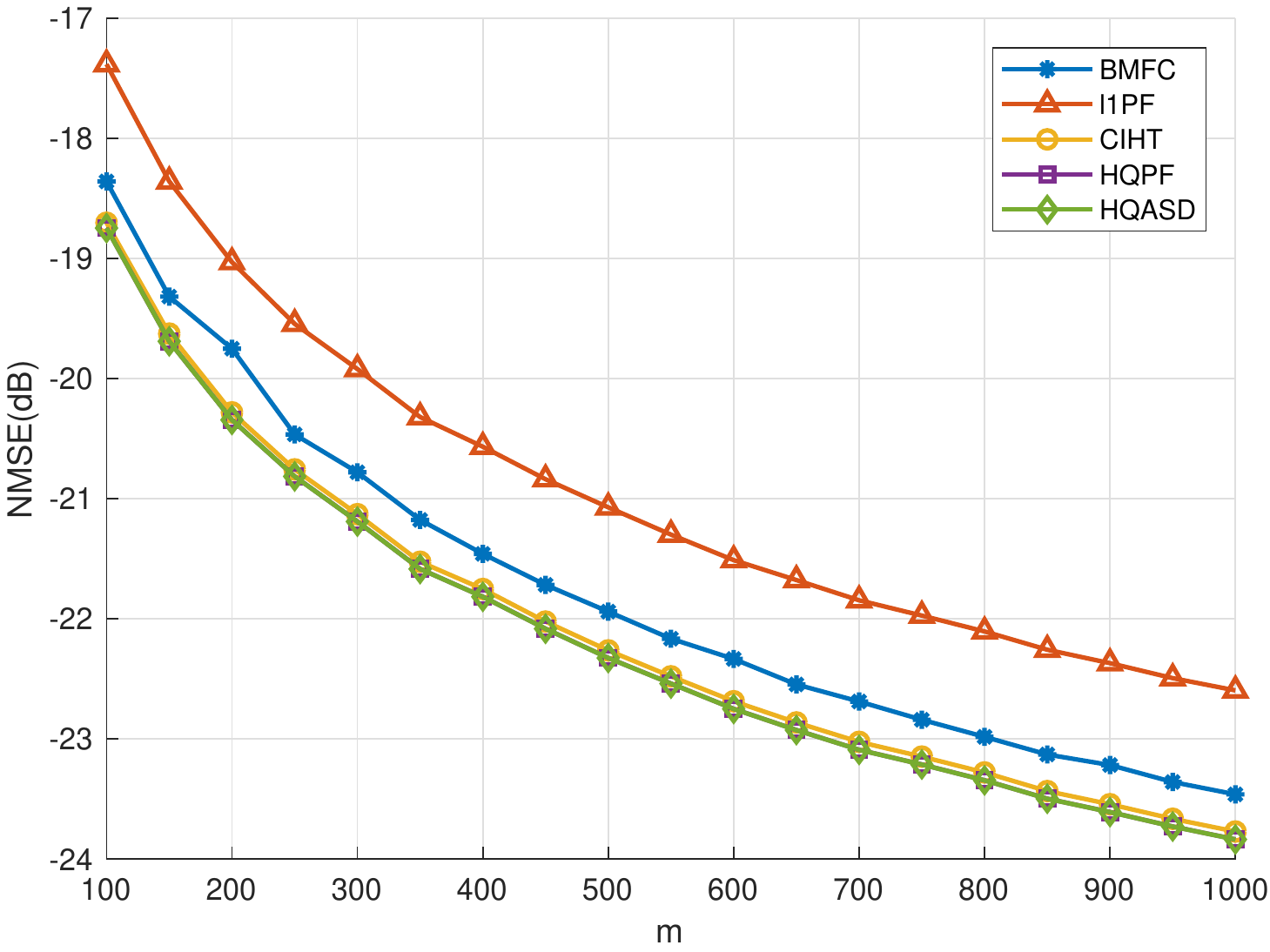}
\caption{Curves of NMSE under different matrix sizes $m$.}
\label{fig5}
\end{figure}

\begin{figure}[tb]
\centering
\includegraphics[width=0.9\linewidth]{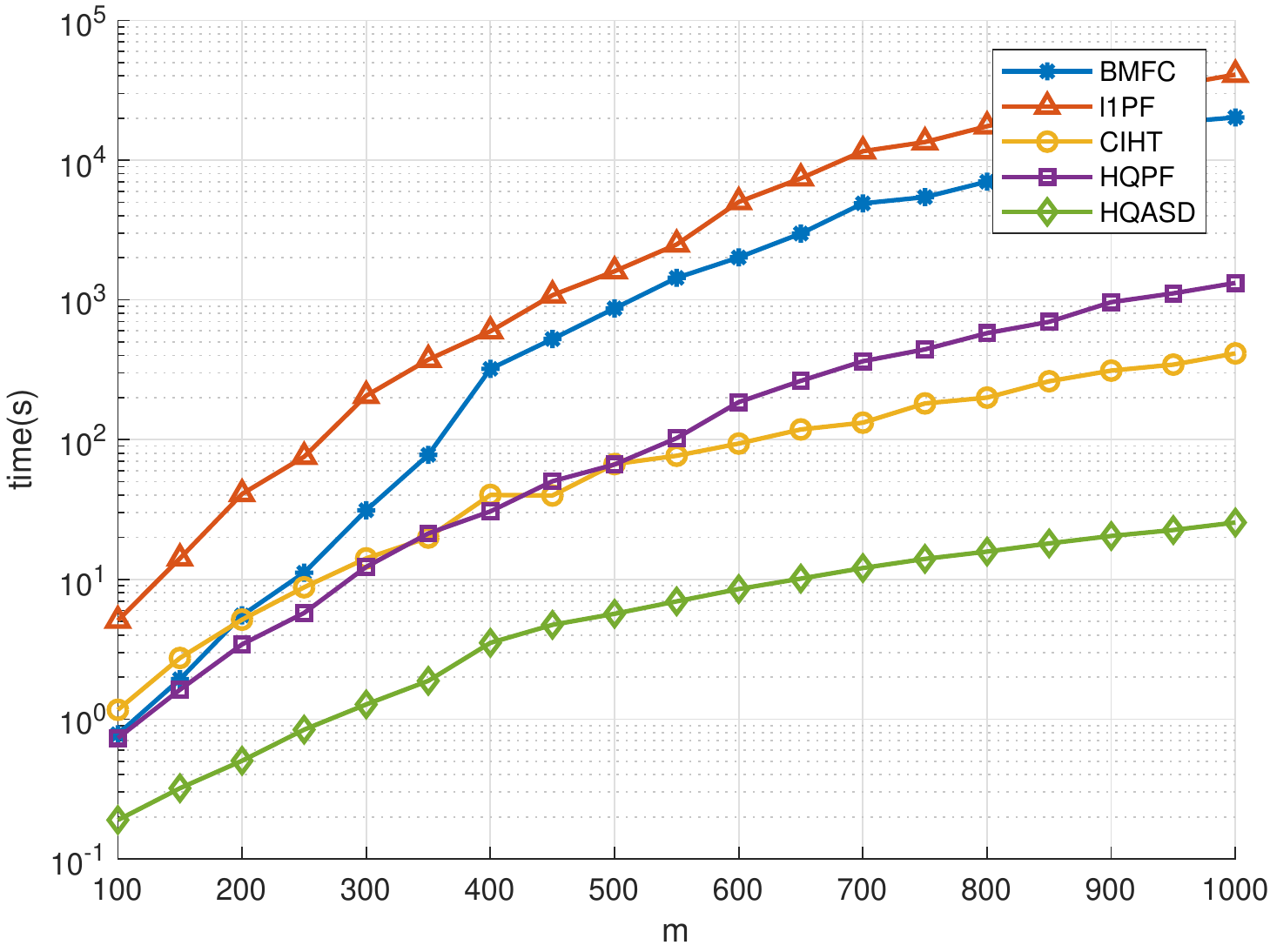}
\caption{Curves of average running times versus matrix size $m$.}
\label{fig6}
\end{figure}

 Second, the performance with different sizes of $\boldsymbol{X}$ are investigated. Fig.\ref{fig5} shows the curves of NMSE in terms of different matrix sizes $m$, and Fig.\ref{fig6} shows the corresponding average running times for a single matrix completion procedure. As can be seen, as the size of the matrix increases, the NMSE values of all algorithms decrease, while the time costs increase significantly for all algorithms. The proposed two algorithms HQPF and HQASD achieve comparable lower NMSE than other algorithms, and HQASD runs much faster than other algorithms. In particular, HQASD can run as fast as about 3 orders of magnitude than $l_1$-PF and BMFC when the matrix size is larger than 700.
\par
\begin{figure*}[tb]
\centering
\includegraphics[width=0.98\linewidth]{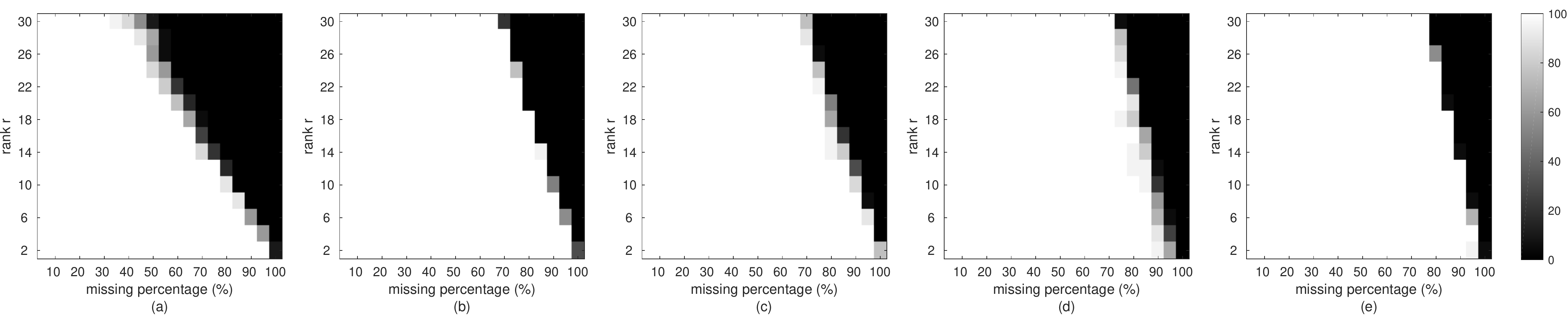}
\caption{Phase transition image for different algorithms. (a) BMFC, (b) $l_1$-PF, (c) CIHT, (d) HQPF, (e) HQASD.}
\label{fig7}
\end{figure*}

\begin{figure}[tb]
\centering
\includegraphics[width=0.9\linewidth]{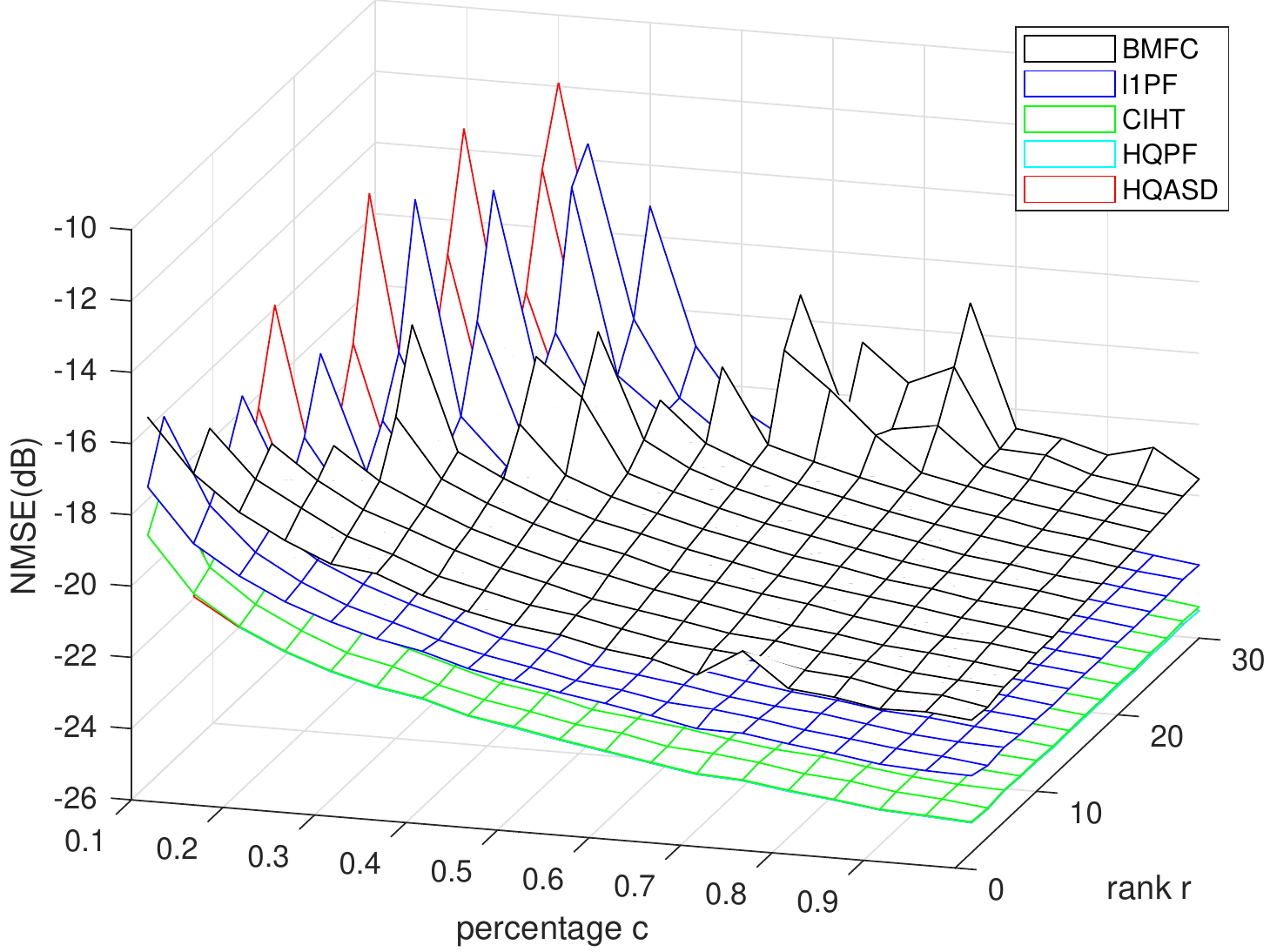}
\caption{Surfaces of NMSE with different rank $r$ and observation percentage $p$.}
\label{fig8}
\end{figure}

\begin{figure}[tb]
\centering
\includegraphics[width=0.9\linewidth]{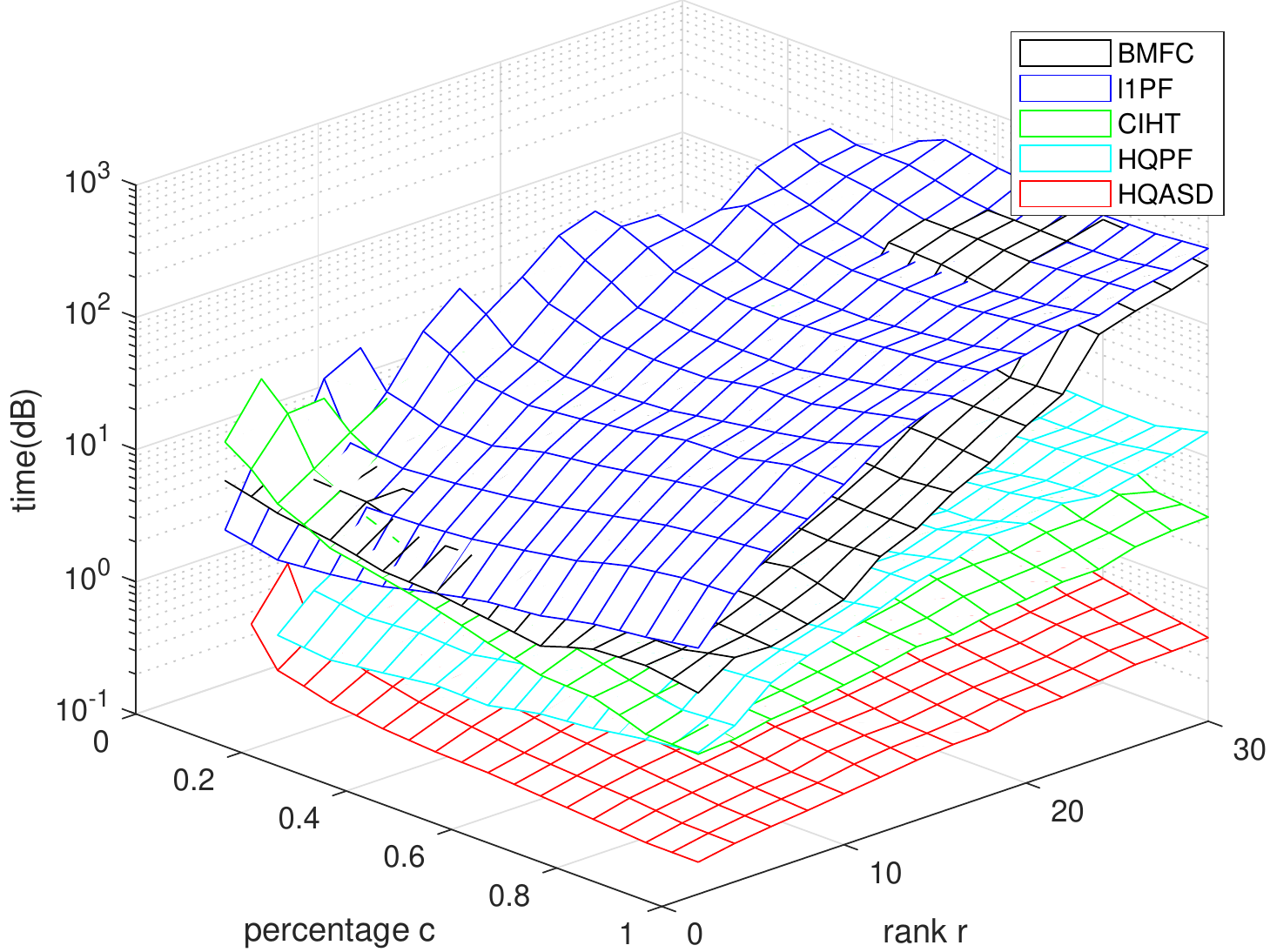}
\caption{Surfaces of average running times with different rank $r$ and observation percentage $p$.}
\label{fig9}
\end{figure}

 Third, we explore the largest recoverable rank of $\boldsymbol{X}$ under different percentage of observed entries $p$, which is also called the phase transition. The rank $r$ and the percentage $p$ are set within [2,30] and [0,100], respectively. For each selection of $r$ and $p$, 200 Monte Carlo runs are performed, and the recovery is judged to be successful if the NMSE is lower than $10e^{-1}$. The phase transition image for different algorithms is shown in Fig.\ref{fig7}. The shade of the color block represents the probability of success, i.e. the percentage of successful recovery in 200 Monte Carlo runs. One can observe that the white region of HQASD and HQPF is larger than other algorithms, which shows that the proposed algorithms can afford larger rank or lower percentage of observations. Meanwhile, the corresponding average running times and NMSE values curves for each algorithm are shown in Fig.\ref{fig8} and Fig.\ref{fig9}. Note that only corresponding data with white blocks in Fig.\ref{fig7} are shown in Fig.\ref{fig8} and Fig.\ref{fig9}. As seen, the proposed HQASD achieve lowest NMSE as well as lowest computational cost among all algorithms. HQPF and CIHT achieve similar low NMSE with HQASD but need more time for completion. Moreover, among all distributable algorithms, the HQPF performs the best.
\par
\begin{figure}[tb]
\centering
\includegraphics[width=0.9\linewidth]{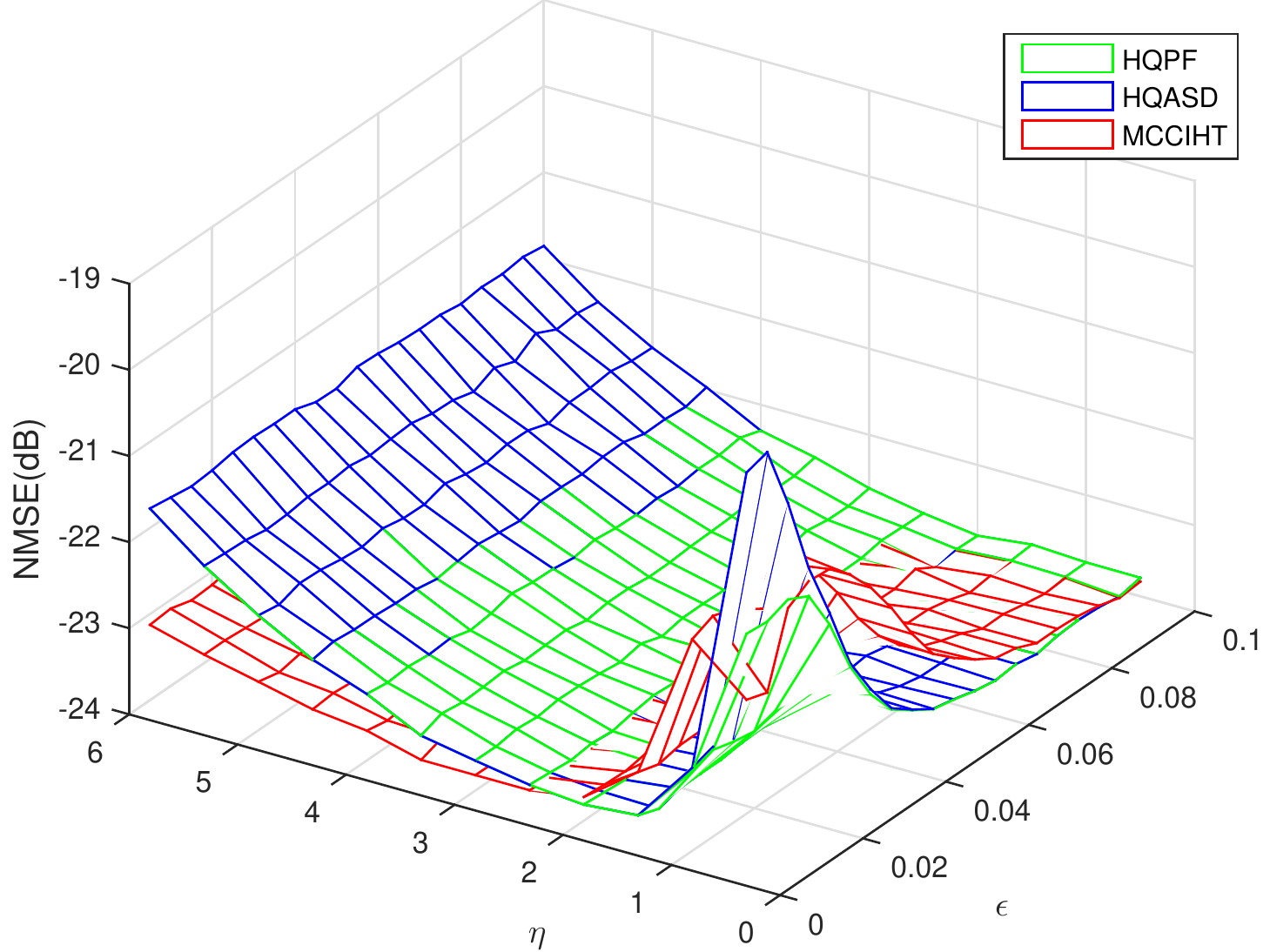}
\caption{Surfaces of NMSE with different $\eta$ and $\xi$.}
\label{fig10}
\end{figure}

 It is known that for correntropy based algorithms, the kernel width highly affects the performance. Thus, here we also analyze the sensitivity of kernel parameters. As shown in Eq.(\ref{sigmaout}), in kernel adaptation strategy, the kernel width $\sigma$ is determined by the choice of $\eta$ and $\xi$. Thus we select different values of $\eta$ and $\xi$ and then obtain the corresponding NMSE over 200 Monte Carlo runs. The NMSE curves versus different $\eta$ are shown in Fig.\ref{fig10}. As can be seen, the algorithms can work well in a wide range of values of $\eta$ and $\xi$. In particular, when $\eta>2$, the NMSE increases as $\eta$ grows, while the NMSE will not highly be affected when selecting different values of $\xi$. When the $\eta$ and $\xi$ are both selected as the too small values, the kernel width $\sigma$ is too small and the algorithms may not converge to local minima.

\subsection{Image Inpainting}
In this part, we compare the performance of algorithms on image inpainting tasks with non-Gaussian noise. Image inpainting aims to fill in the unknown pixels of an image from an incomplete image. Because the most image can be well approximated by low-rank representation, image inpainting can be seen as a matrix completion task. Moreover, to evaluate the performance under non-Gaussian noise, we select a mixture of Gaussian and Salt-and-pepper noise as the noise model. Gaussian noise is a typical normal image noise caused by electronic components. Salt-and-pepper noise is another noise produced in errors in the analog-to-digital converter or bit transmission caused by sudden intense interference, such that the noise value with $0$ or $1$ sparsely occurs on the image. We utilize the popular peak signal to noise ratio (PSNR) to evaluate the performance, which is defined as
$$PSNR=\frac{nm}{{{\| {{\hat {\boldsymbol{M}}}}-\boldsymbol{X} \|}_F^2}}$$
A higher PSNR represents better recovery performance.

\begin{figure*}[tb]
\centering
\includegraphics[width=0.8\linewidth]{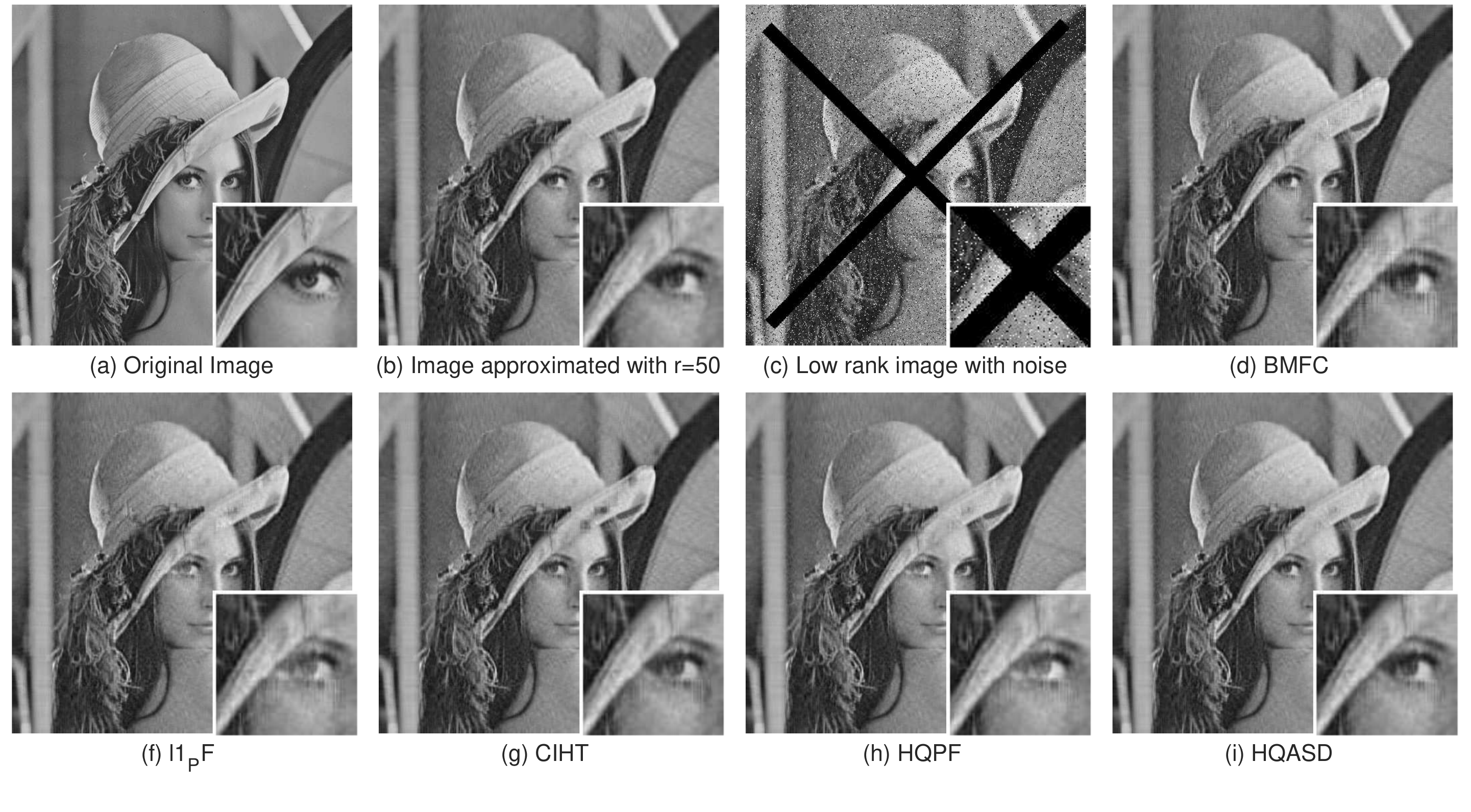}
\caption{Image inpainting sample of Lena image under mixture of Gaussian and Salt-and-pepper noise.}
\label{fig11}
\end{figure*}

\begin{figure*}[tb]
\centering
\includegraphics[width=0.8\linewidth]{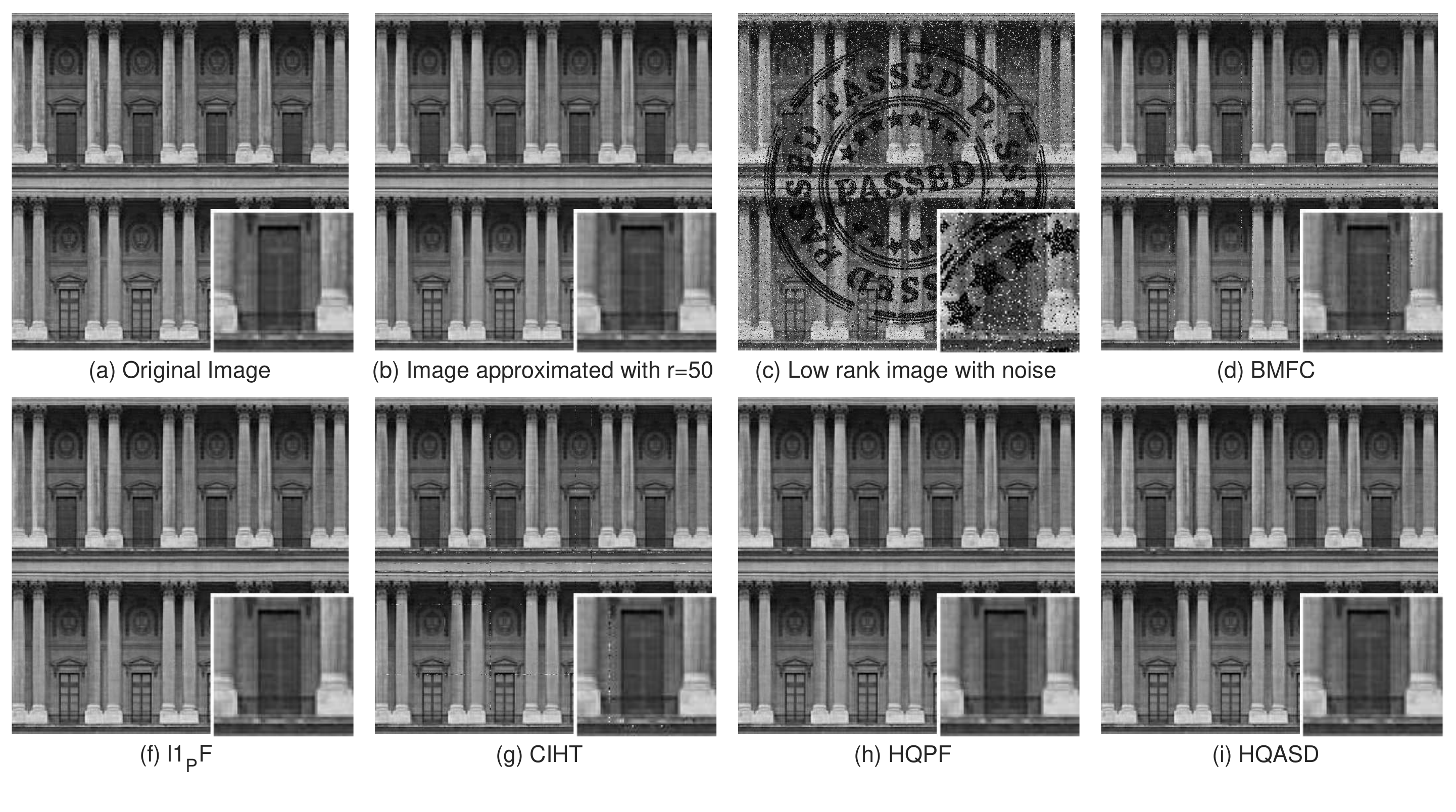}
\caption{Image inpainting sample of Palace image under mixture of Gaussian and Salt-and-pepper noise.}
\label{fig12}
\end{figure*}

We select the $512\times512$ Lena and Palace image as the test images. Lena image (Fig.\ref{fig11}(a)) is a popular image for performance evaluation, while Palace image (fig.\ref{fig12}(a)) contains duplicate patterns which is always utilized for image inpainting test. Each image is compressed via best rank-50 approximation (see. Fig.\ref{fig11}(b) and Fig.\ref{fig12}(b)) so that the low rank property is guaranteed. Then the two images are masked in a "cross pattern" and a "stamp mark", respectively. Finally, the observed pixels of images are added with Gaussian noise with variance $0.0001$, and then $10\%$ of the observed pixels are also disturbed by salt-and-pepper noise. During the simulations, 100 Monte Carlo runs are performed for each recovery task. The algorithms parameters for HQPF an HQASD are tuned as $\varepsilon_1=10000$. For BMFC, to obtain better performance, the parameter $a$ is set to 6 and 20 for Gaussian and non-Gaussian noise, respectively.

Table.1 lists the average recovery PSNR and the corresponding average running times under noiseless and noisy environments. One can see that the proposed HQASD algorithm achieves the best performance in all simulations. In particular, HQASD obtains the highest PSNR as well as lowest running times. Moreover, HQPF obtains the comparable PSNR for Palace image, while the performance is worse than HQASD for Lena image. Nevertheless, HQPF still achieves the best performance among distributable algorithms.

To further demonstrate the recovery performance, samples of images recovered by different algorithms under non-Gaussian noise are shown in Fig.\ref{fig11} and Fig.\ref{fig12}. The enlarge view of part of recovered images are also depicted to evidently show the recovery differences. One can see that when filling missing entries of the face region in Lena image, fringes are produced in all of the recovered images. In particular, BMFC and MCCIHT have the most obvious fringes, while HQASD has the least visible fringes. Moreover, $l_1$-PF and HQPF fail to accurately recover the left eye, which is probably caused by converging to a wrong local minimum perturbed by non-Gaussian noise via alternative minimization. Furthermore, for the palace image, the recovered image of BMFC and MCCIHT still have visible reconstruction error. $l_1$-PF, HQPF and HQASD can successfully recover the image. From the enlarged view, one can see that the recovered image of HQPF and HQASD are slightly clear than $l_1$-PF, especially for object edges.

\renewcommand{\arraystretch}{1.5} 
\newcolumntype{C}[1]{>{\centering\arraybackslash}p{#1}}
\begin{table*}[tp]
  \centering
  \fontsize{8}{8}\selectfont
  \begin{threeparttable}
  \caption{Image Inpainting Performance Comparison: PSNR and average running times}
  \label{imagecomparison}
    \begin{tabular}{C{1.8cm}C{1cm}C{1cm}C{1cm}C{1cm}C{1cm}C{1cm}C{1cm}C{1cm}}
    \toprule
    \multirow{3}{*}{Method}&
    \multicolumn{2}{c}{Lena+low rank}&\multicolumn{2}{c}{Lena+low rank+noise}&\multicolumn{2}{c}{Palce+low rank}&\multicolumn{2}{c}{Palce+low rank+noise}\cr
    \cmidrule(lr){2-3} \cmidrule(lr){4-5} \cmidrule(lr){6-7} \cmidrule(lr){8-9}
    &PSNR(dB)&Time(s)&PSNR(dB)&Time(s)&PSNR(dB)&Time(s)&PSNR(dB)&Time(s)\cr
    \midrule
    BMFC&13.81& 1459.8& 35.18& 1374.3& 18.71& 706.3& 22.2& 1331.1\cr
    $l_1$-PF&76.90& 1752.9& 34.65& 2575.3& 87.59& 568.3& 42.5& 1839.7\cr
    CIHT&60.33& 188.8& 32.81& 62.58& 26.55& 194.23& 31.6& 62.2\cr
    HQPF&75.33& 465.2& 36.93& 181.34& 88.60& 77.8& {\bf45.3}& 130.3\cr
    HQASD&{\bf 92.88}& {\bf27.5}& {\bf42.44}& {\bf 5.75}& {\bf88.84}& {\bf3.7}&{\bf 45.3}& {\bf3.6}\cr
    \bottomrule
    \end{tabular}
    \end{threeparttable}
\end{table*}

\subsection{Experiments on MovieLens dataset}
In this part we evaluate the proposed algorithm on the real data set. MovieLens is a widely used dataset for recommender system. Firstly, similar to experiments in \cite{Zhao2016Efficient,Zeng2018Outlier}, we carry out the experiment on MovieLens-100K data set. MovieLens-100K consists of 100,000 ratings (1-5) from 943 users on 1682 movies, and the percentage is observed data about $6\%$. It also provide 5 splits of training data $\boldsymbol{X}_{train}$ and testing data $\boldsymbol{X}_{test}$, where $\boldsymbol{X}_{train}$ and $\boldsymbol{X}_{test}$ account for $80\%$ and $20\%$ of the observed data, respectively. We perform the test on both noiseless case and noisy case. In noisy case, $10\%$ of the rating value $1$ are replaced by $5$, and $10\%$ of the rating value $5$ are replaced by $1$, too. The performance is evaluated using the root mean square error (RMSE) defined as \cite{Zhao2016Efficient}
$$RMSE=\sqrt{\frac{{{E\left[ {\left\|\boldsymbol{\Omega}_{test}\circ\left({\hat {\boldsymbol{M}}}-\boldsymbol{X}_{test}\right) \right\|}_F^2\right]}}}{{\rm{card}}(\boldsymbol{\Omega}_{test})}}$$
where $\boldsymbol{\Omega}\in\mathbb{R}^{m\times n}$ is a logical matrix for testing data where the each entry ${\Omega}_{i,j}\in{0,1}$ denotes whether the ${i,j}$-th entry of testing data $\boldsymbol{X}_{test}$ is observed. The expectation is approximated by 10 Monte Carlo realizations.

\begin{figure}[tb]
\centering
\includegraphics[width=0.9\linewidth]{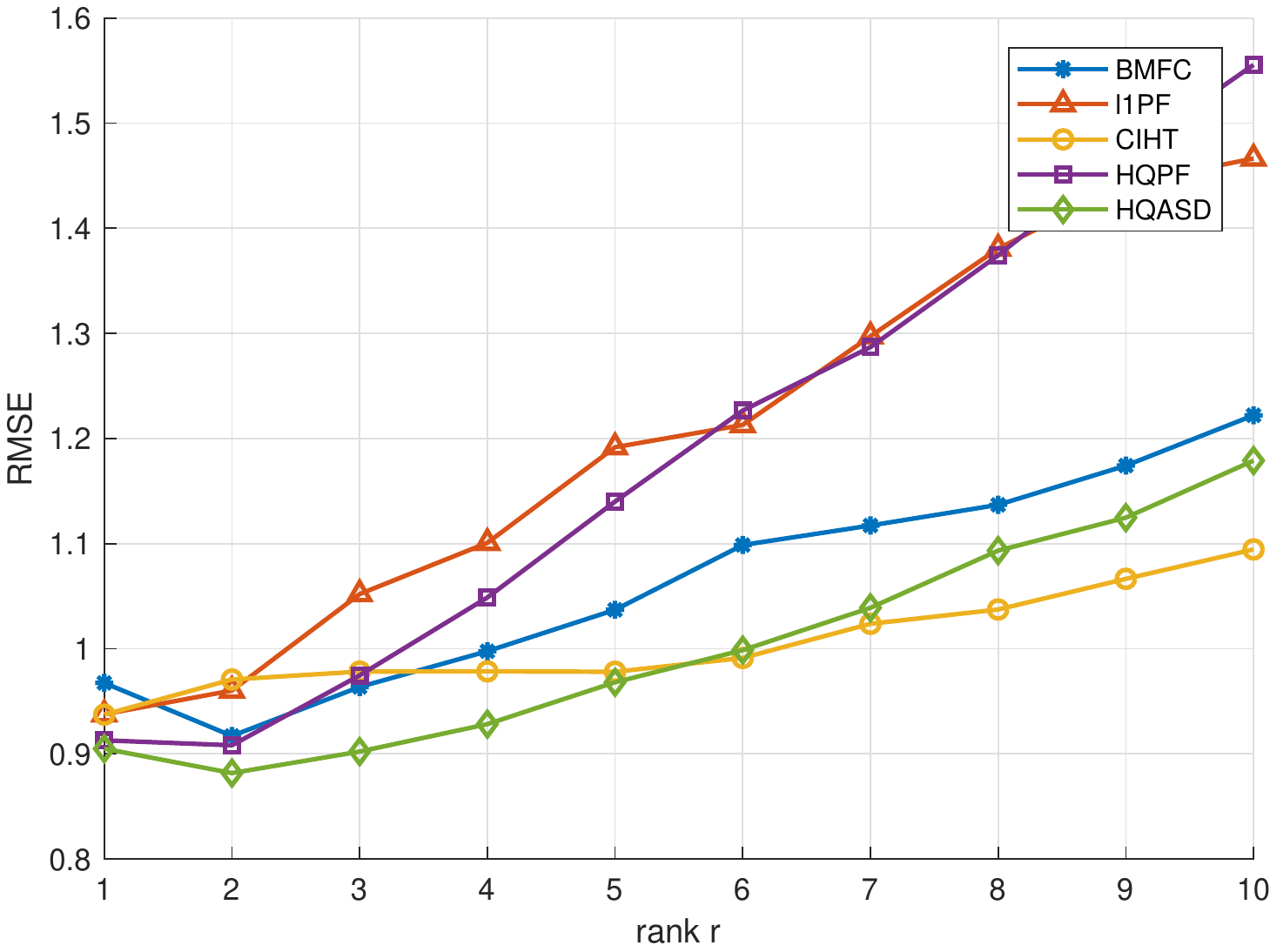}
\caption{RMSE curves with different rank $r$ without noise.}
\label{fig13}
\end{figure}

\begin{figure}[tb]
\centering
\includegraphics[width=0.9\linewidth]{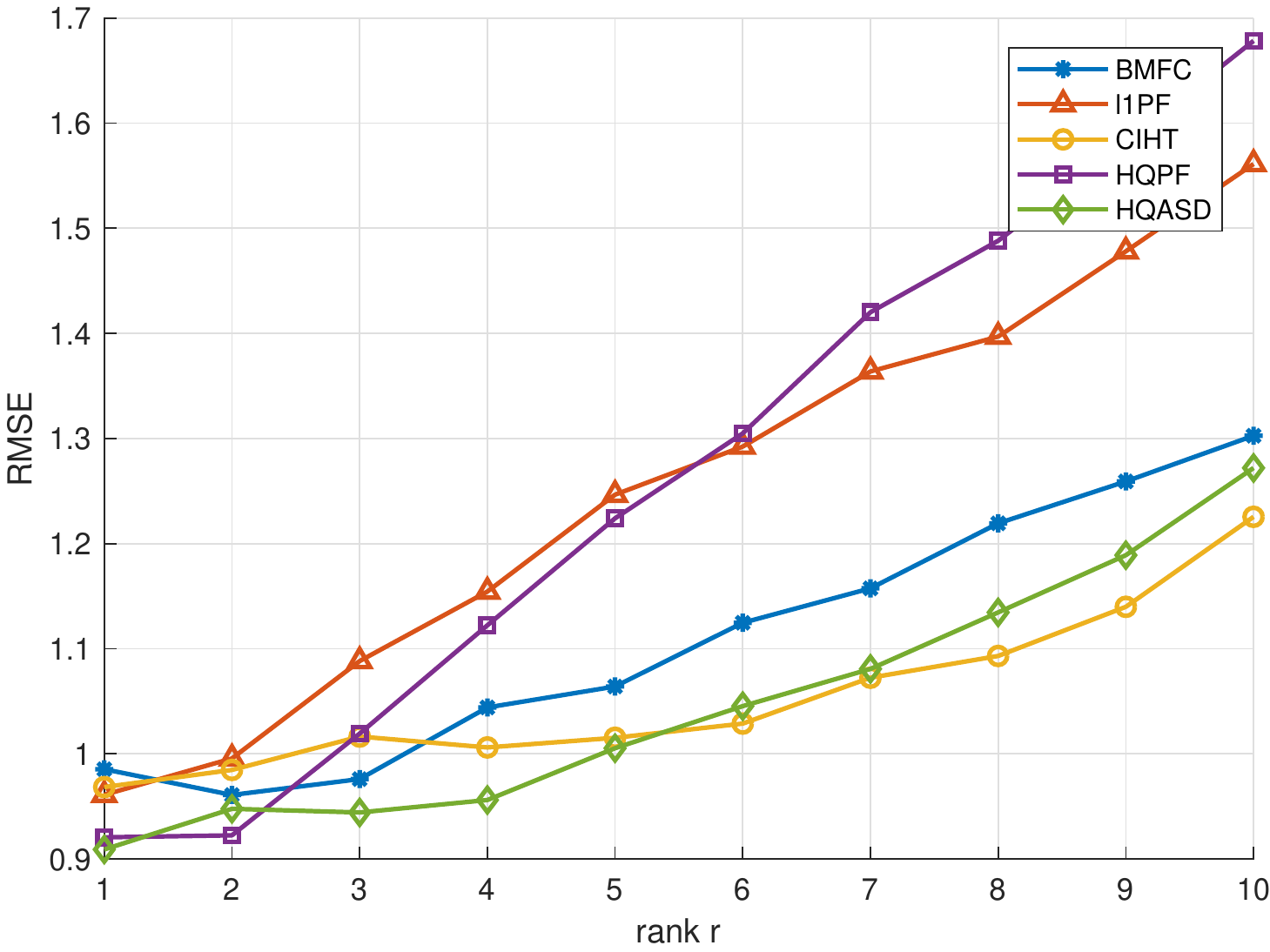}
\caption{RMSE curves with different rank $r$ under noisy environment.}
\label{fig14}
\end{figure}

In this experiment, we set $\varepsilon_1=10000$ for HQPF and HQASD. For CIHT and HQASD, the threshold $\varepsilon_2$ is set to $10e^{-3}$. Fig.\ref{fig13} and Fig.\ref{fig14} depict the RMSE results for all algorithms with different values of rank $r$ under the noiseless and noisy case, respectively. As seen, all algorithms work well when $r=1$ or $2$ in both noiseless and noisy case. While when $r>2$, all algorithms suffer from different degrees of performance degradation. In particular, HQASD can maintain good performance when $r$ is as large as 5. On the contrary, the performance of alternative minimization based $l_1$-PF and HQPF algorithms degrades seriously when $r>2$. Furthermore, we list the average RMSE and corresponding average running times under $r=2$ in Table II. Here we also add traditional $l_2$-based PF algorithm for comparison. One can observe that the HQASD achieves the lowest RMSE in both noiseless and noisy cases. HQASD also runs much faster than other robust methods.

To further demonstrate the advantage of proposed algorithms in computational cost, we also carry out the comparison on the more challenging MovieLens-1m dataset. The MovieLens-1m dataset contains 1,000,209 anonymous ratings of approximately 3,900 movies made by 6,040 users. The observation percentage is only 4$\%$, and the matrix size is 15 times larger than MovieLens-100K. We evaluate the performance of all algorithms on MovieLens-1m under noiseless and noisy cases similar to experiments on MovieLens-100K. The noise and algorithm settings are the same as the previous simulation. Table II shows the average RMSE and corresponding average running times. One can observe that the cost times increase greatly compared with MovieLens-100K for all algorithms. The RMSE also increases compare to MovieLens-100K. In particular, HQASD still achieves much more fast performance than other robust algorithms, and HQPF obtains the best performance among distributable algorithms.

\renewcommand{\arraystretch}{1.5} 
\newcolumntype{C}[1]{>{\centering\arraybackslash}p{#1}}
\begin{table*}[tp]
  \centering
  \fontsize{8}{8}\selectfont
  \begin{threeparttable}
  \caption{MovieLens Dataset Performance Comparison: RMSE and average running times}
  \label{imagecomparison}
    \begin{tabular}{C{1.8cm}C{1cm}C{1cm}C{1cm}C{1cm}C{1cm}C{1cm}C{1cm}C{1cm}}
    \toprule
    \multirow{3}{*}{Method}&
    \multicolumn{2}{c}{MovieLens100K}&\multicolumn{2}{c}{MovieLens100K+noise}&\multicolumn{2}{c}{MovieLens1M}&\multicolumn{2}{c}{MovieLens1M+noise}\cr
    \cmidrule(lr){2-3} \cmidrule(lr){4-5} \cmidrule(lr){6-7} \cmidrule(lr){8-9}
    &RMSE&Time(s)&RMSE&Time(s)&RMSE&Time(s)&RMSE&Time(s)\cr
    \midrule
    $l_2$-PF&0.9494& {\bf9.1} & 0.9776& 18.6& 1.1419& {\bf162.7}& 1.1457& {\bf122.4}\cr
    BMFC&0.9849& 104.2& 1.0080& 118.7& 1.1516& 2185.7& 1.1751& 2042.3\cr
    $l_1$-PF&0.9912& 242.4& 1.0009& 213.6& 1.1955& 5734.8& 1.2045& 5024.1\cr
    CIHT&0.9916& 434.4& 1.0012& 412.8& 1.1411& 18694.0& 1.1417& 17295.2\cr
    HQPF&0.9506& 42.4& 0.9745& 64.6& 1.1441& 3058.3& 1.1437& 2550.0\cr
    HQASD&{\bf 0.9395}& 25.6& {\bf 0.9541}& {\bf 18.5}& {\bf1.1342}& {445.5}&{\bf 1.1356}& 358.7\cr
    \bottomrule
    \end{tabular}
    \end{threeparttable}
\end{table*}

\section{Conclusion}
In this paper, we proposed two novel efficient and robust matrix completion algorithms. The algorithms apply correntropy as the error measure to improve the robustness. To overcome the complicated computation of non-quadratic correntropy based optimization, we utilize the half-quadratic technique to efficiently solve the problem. The two proposed algorithms, HQPF and HQASD, adopt the same half-quadratic method but are developed in different ways. HQPF is derived from traditional alternating minimization method and can be processed in parallel. HQASD is obtained by gradient descend method and has much lower computational cost. Further, an adaptive kernel width selection strategy is proposed to speed up the convergence of the new algorithms. Extensive simulations and real-world data experiments are conducted, demonstrating that the proposed algorithms can achieve better performance than existing state-of-the-art methods.

\section*{Acknowledgements}
This work was supported by 973 Program (No.2015CB351703), National NSF of China (No.61273366 and No.91648208) and National Science and Technology support program of China (No.2015BAH31F01).

\bibliographystyle{IEEEtran}
\bibliography{HQ-MF}

\end{document}